\documentclass[11pt]{article}
\usepackage{amsmath,amssymb,amsfonts}
\usepackage{latexsym}
\usepackage{pstricks,pst-node,pst-text,pst-3d}
\input  epsf 

\newcounter{mycnt}
\def\themycnt{\thesection.\arabic{mycnt}}
\def\mybenv#1{\refstepcounter{mycnt}%
       \vskip 3pt\noindent{\bf #1~~\themycnt}:~}
\def\myeenv{\hfill\rule{1ex}{1ex}\vskip 3pt}

\def\abc{{[}{}^{}_a{}^b_c{]}}
\def\cab{{[}{}^{\mbox{}}_c{}^a_b{]}}
\def\bac{{[}{}^{\mbox{}}_b{}^a_c{]}}
\def\abcd{{[}{}^a_b{}^c_d{]}}
\begin{document}
\begin{titlepage}
\begin{flushright}
UTAS-PHYS-2010-08\\
October 2010, Revised March 2011\\
\end{flushright}
\begin{centering}
 
{\ }\vspace{0.5cm}
 
{\Large\bf A class of quadratic deformations of Lie superalgebras\footnote{Original title: Finite dimensional quadratic Lie superalgebras}}

%

\vspace{1cm}
P. D. Jarvis\footnote{{\tt Peter.Jarvis@utas.edu.au}}\\
{\em School of Mathematics and Physics, University of Tasmania},
\\[.3cm]
G. Rudolph\footnote{{\tt rudolph@rz.uni-leipzig.de}}\\
{\em Institute for Theoretical Physics, University of Leipzig},\\[.3cm]
and 
L. A. Yates\\
{\em School of Mathematics and Physics, University of Tasmania}.\\

\vspace{1cm}

\thispagestyle{empty}
\begin{abstract}
\noindent
We study certain ${\mathbb Z}_2$-graded, finite-dimensional polynomial algebras of degree 2 which are a special class of deformations of  Lie superalgebras, which we call quadratic Lie superalgebras. Starting from the formal definition, we discuss the generalised Jacobi relations in the context of the Koszul property, and give a proof of the PBW basis theorem. We give several concrete examples of quadratic Lie superalgebras for low dimensional cases, and discuss aspects of their structure constants for the `type I' class. We derive the equivalent of the Kac module construction for typical and atypical modules, and a related direct construction of irreducible modules due to Gould. We investigate in detail one specific case, the quadratic generalisation $gl_2(n/1)$ of the Lie superalgebra $sl(n/1)$. We formulate the general atypicality conditions at level 1, and present an analysis of zero-and one-step atypical modules for a certain family of Kac modules. 
\end{abstract}

\noindent

\vspace{10pt}

\end{centering} 

\vspace{125pt}

\end{titlepage}
\tableofcontents
\setcounter{footnote}{0}
\section{Introduction}
\label{sec:Introduction}
The classical problem of investigation of the correct mathematical structures within which to frame physical models is, in the case of quantum mechanical systems, intimately related to underlying symmetry principles. Until recently, the classical mathematics of discrete groups and Lie groups and their associated algebras and representations were the framework within which regularities of energy levels and matrix elements were organised, even in the absence of microscopic theories. The role of such underlying symmetry principles has not been supplanted even with the advent of supersymmetry and Lie superalgebras to handle fermionic states.
The revolutions of quantum groups and integrable models in one dimensional quantum field theory or two dimensional lattice models have shown in recent decades that more subtle mathematical structures, such as braiding for tensor products of composite state vector spaces, can and should have a natural place in the physicist's tool-kit. These objects, centring around the famous Yang-Baxter equation, are definitively `nonlinear' in their structure, and often entail one or more additional `deformation' parameters which can be new physical constants, presaging the breakdown of standard symmetry relations in the appropriate asymptotic regime.


The present work deals with some classes of \emph{polynomial} algebras. These are mild generalisations of Lie algebras, but are surprisingly universal, in that they find applications
in many different fields, and also contain special cases of some of the above-mentioned quantum groups.
Our study is restricted to certain polynomial, ${\mathbb Z}_2$-graded algebras of degree 2, 
which are a special class of deformations of  Lie superalgebas, which we term `quadratic Lie superalgebras'. In our previous work, we studied these objects in the context of the problem of enumerating gauge invariant fields (observables) in a Hamiltonian lattice formulation of QCD. In \cite{JarvisRudolph2003} we introduced and studied `polynomial super $gl(n)$ algebras'  heuristically via free-fermion and -boson constructions.  The methodology resembled the intention of very early attempts to define `baryonic currents' for the purpose of generalising the applicability of scalar and vector meson current algebras in particle physics \cite{DelbourgoSalamStrathdee1966mfr}. In
\cite{JarvisKijowskiRudolph2005} our constructions were used to study the structure of the 
observable algebra of lattice QCD and to prove a classification theorem for 
its irreducible representations (superselection structure).

In the present work we are at pains to establish a systematic approach to quadratic Lie superalgebras in their own right. They are introduced in \S \ref{sec:FormalPBW} by way of formal definitions via appropriate tensor algebras and defining relations. For completeness we refer to the abstract context of Koszul complexes and Koszulness. Although the latter attribute is needed, it is not further explored in this paper as our notation is anchored in concretely enumerated structure relations.
We cite the important structure theorem (Theorem \ref{thm:KoszulQuadSuperalg}) establishing the existence and significance of generalised Jacobi identities, which if true together with Koszulness, establishes the existence of the PBW basis theorem. In Lemma  \ref{lem:PBWbasis} we give a direct proof of the PBW property (detailed in the appendix, \S \ref{subsec:PBWproof}) which guarantees Koszulness. In \S  \ref{sec:TypeIKacGould}, we introduce a natural specialisation to a sub-class of quadratic Lie superalgebras, the so-called type I class, which admits a refined grading with a specific structure. Although it is beyond the scope of our present techniques to formulate a classification of quadratic Lie superalgebras, nonetheless some features of the structure constants are generic, with an intimate relation to invariants of the even subalgebra, as shown by Lemma \ref{lem:StructureConsts} and its proof. 
The results of \S \ref{sec:FormalPBW} are used together with the grading to establish useful alternative decompositions of the universal enveloping algebra. In turn, this sets the scene in \S \ref{sec:TypeIKacGould} for the definition of induced representations following the Kac module \cite{Kac1977lsa,Kac1978rcls} construction in the Lie superalgebra case, and as modified by Gould \cite{Gould1989ar} to give a presentation of all finite dimensional irreducible Kac modules, including those of atypical type.
In \S \ref{sec:Atypicals} we turn to a specific family of quadratic Lie superalgebras which we identified previously, the quadratic generalisations $gl_2(n/1)$ of the classical Lie superalgebra $sl(n/1)$. Using the methods of characteristic identities for generators of classical groups \cite{green1971ci,BrackenGreen1971vo} as applied to the Lie superalgebra case 
\cite{Gould1989ar,GouldBrackenHughes1989br} we provide an explicit characterisation of atypicality conditions for irreducible modules at level one. For a specific class of highest weight modules we also examine zero- and one-step atypicals, and give some explicit cases. In the concluding \S \ref{sec:Examples} we recall some of the previously-identified quadratic Lie superalgebras motivated from lattice QCD, and note how the phenomenon of atypicality arises there. One of the features of our analysis, both here and in \S \ref{sec:Atypicals}, is the remarkable fact
that type I quadratic Lie superalgebras generically admit atypical modules of depth \emph{zero} -- that is, they are representable on a \emph{single}, nontrivial irreducible module of the even subalgebra.


\section{Formal constructions and the PBW basis theorem}
\label{sec:FormalPBW}

Let $L= L_{\bar{0}} + L_{\bar{1}}$ be a finite-dimensional ${\mathbb Z}_2$-graded complex vector space.
Take the even and odd subspaces $L_{\bar{0}}$ and $L_{\bar{1}}$ to be spanned by basis elements $x_i$, $i=1,2,\cdots,n$, and $y_r$, $r=1,2,\cdots,m$, respectively.  The tensor algebra $T(L) = \sum_{n=0}^\infty \otimes^n(L)$ $\cong {\mathbb C} + L + L\otimes L + \cdots$ inherits the ${\mathbb Z}_2$-grading in the natural way, in that for $T^n := \otimes^n T(L)$ we have
\begin{align}
\left. T^n \right|_{\bar{0}} \cong  & \,  {\sum}_{\sum \bar{\alpha}_i \equiv \bar{0}} L_{\bar{\alpha}_1} \otimes L_{\bar{\alpha}_2} \otimes \cdots  \otimes L_{\bar{\alpha}_n}, \qquad
\left. T^n \right|_{\bar{1}} \cong  {\sum}_{\sum \bar{\alpha}_i \equiv \bar{1}}  L_{\bar{\alpha}_1} \otimes L_{\bar{\alpha}_2} \otimes \cdots  \otimes L_{\bar{\alpha}_n}, \nonumber
\end{align}
with each $\bar{\alpha}_i =\bar{0}$ or $\bar{1}$, and $T^n \cong \left. T^n \right|_{\bar{0}}+\left. T^n \right |_{\bar{1}}$.

We now introduce the concrete defining relations of the graded algebras of interest.
The explicit Definition \ref{def:QuadSuperalg} below will be justified by a discussion of the structure of the enveloping algebra, to be given presently in a slightly more general context, after which the definition and the motivation for taking up this class as a natural object of investigation will become clearer (see Remarks \ref{comm:Explanations} below).

Consider the following arrays of complex numbers: ${d_{p,q}}^{k,\ell} = {d_{q,p}}^{k,\ell} ={d_{p,q}}^{\ell, k} $; 
${b_{p,q}}^k =  {b_{q,p}}^k$; ${a_{p,q}}=a_{q,p}$, as well as ${c_{i,j}}^k = - {c_{j,i}}^k$, and $\bar{c}_{k,p}{}^q$, for indices in the specified ranges above, namely $i,j,k,\ell = 1,2,\cdots,n$, and $p,q,r = 1,2,\cdots,m$.  They are required to satisfy the following bilinear relations (summation on repeated indices, and for ease of writing, the comma ${}_{-,-}$ separating suffixes is omitted where there is no ambiguity):
\begin{align}
{c_{ij}}^\ell{c_{\ell k}}^m=& \, {c_{ik}}^\ell{c_{\ell j}}^m+{c_{jk}}^\ell{c_{\ell i}}^m ; \nonumber \\
{c_{ij}}^\ell \bar{c}_{\ell p}{}^q = & \, \bar{c}_{ir}{}^q\bar{c}_{jp}{}^r-\bar{c}_{jr}{}^q\bar{c}_{ip}{}^r ; \nonumber \\
{c_{in}}^k {d_{pq}}^{n\ell}+ {c_{in}}^\ell {d_{pq}}^{k n} = & \,
\bar{c}_{ip}{}^s {d_{sq}}^{k\ell} + \bar{c}_{ip}{}^s {d_{sq}}^{k\ell} ,
\nonumber \\
{b_{pq}}^m {c_{im}}^n = & \, \bar{c}_{ip}{}^s {b_{sq}}^n + \bar{c}_{iq}{}^s {b_{ps}}^n;
\nonumber \\
\bar{c}_{mp}{}^s {b_{qr}}^m + \bar{c}_{mq}{}^s {b_{rp}}^m + \bar{c}_{mr}{}^s {b_{pq}}^m = & \, 0,
\nonumber \\
\bar{c}_{mp}{}^s {d_{qr}}^{m\ell} + \bar{c}_{mq}{}^s {d_{rp}}^{m\ell} + \bar{c}_{mr}{}^s {d_{pq}}^{m\ell} = & \, 0.
\label{eq:GenJacobiComponents}
\end{align}
Let $I$ be the linear subspace of $L\otimes L + L + {\mathbb C}$ spanned by the set
\begin{align}
x_i \otimes x_j - x_j \otimes x_i -  {c_{ij}}^k x_k; & \, \nonumber \\
y_p \otimes y_q +y_q\otimes  y_p  -  {d_{pq}}^{k\ell} x_k\otimes  x_\ell - {b_{pq}}^k x_k - {a_{pq}}; & \, \nonumber \\
x_i\otimes  y_p - y_p \otimes x_i - \bar{c}_{ip}{}^q y_q,& \, 
\label{eq:QuadIdeals}
\end{align}
and let $J(I)$ be the corresponding two-sided ideal in $T(L)$ generated by $I$.

\mybenv{Definition} \textbf{Quadratic Lie superalgebra:}\\
\label{def:QuadSuperalg}
Given quantities ${d_{p,q}}^{k,\ell}$, ${b_{p,q}}^k$, ${a_{p,q}}$, ${c_{i,j}}^k$ and $\bar{c}_{k,p}{}^q$ with properties (\ref{eq:GenJacobiComponents}) above, and the corresponding subspace $I$ as in (\ref{eq:QuadIdeals}) and ideal $J(I)$, the subalgebra of the tensor algebra defined by $U(I)=T(L)/{J(I)}$ is called the \emph{quadratic Lie superalgebra} associated with $I$. \myeenv 

\noindent
Associated with the tensor algebra is the filtration defined by $T_n = \sum_{k=0}^n T^k$, in such a way that $ {\mathbb C}\cong T_0 \subset T_1 \subset T_2 \subset \cdots$, that is, $T_n \subset T_{n\!+\!1}$. $U\equiv U(I)$ inherits this filtration in the natural way, so that we have ${\mathbb C}\cong U(I){}_0 \subset  
U(I){}_1 \subset U(I){}_2 \subset \cdots $ and we define the associated graded algebra as the direct sum:

\mybenv{Definition} \textbf{Graded algebra associated to quadratic Lie superalgebra:}\\
\label{def:GradedQuadSuperalg}
\noindent
The  graded algebra $gr(U(I)) ={\mathbb C} + \sum_{n=1}^\infty U(I){}_n/U(I){}_{n\!-\!1}$ is called the \emph{graded superalgebra} associated with the quadratic Lie superalgebra $U(I)$. \myeenv 
\noindent
Where the context is clear, we omit the bracket $(I)$ and simply denote the quadratic Lie superalgebra by $U$, and the associated graded algebra by $gr(U)$.  $I$ is ${\mathbb Z}_2$-graded, and hence so also is the ideal $J(I)$, spanned by elements of $T(L)$ of the form (at degree $p+q+2$, for example) 
\begin{align}
\label{eq:GeneratedSubspaces}
I^p_q \cong & \,  \big( \otimes^{p} L\big) \otimes I \otimes  \big( \otimes^q L\big)
\end{align}
with $p$ prefactors of $L$ and $q$ postfactors. Thus $U$ inherits the natural ${\mathbb Z}_2$ grading from $T(L)$. 

Note that $U$ is strictly speaking a (${\mathbb Z}_2$-graded) \emph{quadratic-linear} algebra in the notation of \cite{PolishchukPositselski2005qa}, because $I \subsetneq L \otimes L$ but contains elements of degree 1. However, just as in the case of {Lie} superalgebras, the appellation quadratic \emph{Lie} super\emph{algebra} which we adopt here suggests an intimate relation with an underlying Lie algebra, which is easily seen from the above definitions as follows. Consider the subspace $I_0 \subset L_{\bar{0}} \otimes L_{\bar{0}} + L_{\bar{0}} + {\mathbb C}$ spanned by the set given just by the first line of (\ref{eq:QuadIdeals}) and the corresponding ideal $J({I}_0)$ in $T(L_{\bar{0}})$. Clearly $U_0 := T(L_{\bar{0}})/J({I}_0) \subset U_{\bar{0}}$ simply by inclusion. On the other hand, the linear space $L_{\bar{0}}$ spanned by $x_i$, subject to the constraint given by the first line of (\ref{eq:GenJacobiComponents}), is indeed a  Lie algebra, with the bracket ${[}x_i,x_j{]} := x_i\otimes x_j - x_j\otimes x_i$ , and $U_0 \cong U(L_{\bar{0}})$ the corresponding universal enveloping algebra. By abuse of notation
we refer to $L_{\bar{0}}$ simply as $L_0$, the underlying Lie algebra of the quadratic Lie superalgebra $U$.

\mybenv{Remarks} \mbox{}\\
\label{comm:Explanations}
As in the Lie superalgebra case, the basis elements $x_i$ are called the even generators, the $y_p$ are the  odd generators, and the arrays ${d_{pq}}^{k\ell}$, 
${b_{pq}}^k$, ${a_{pq}}$, ${c_{ij}}^k$, and $\bar{c}_{kp}{}^q$ are called the structure constants of $U$ (indeed, if ${d_{pq}}^{k\ell}=0$, then $L$ is itself a Lie superalgebra). Note that we do \emph{not} include additional quantities, say ${e_{ij}}^{pq}$, which might provide quadratic modifications to the \emph{first} line of (\ref{eq:QuadIdeals}); nor do we admit additional scalars
$c_{ij}$ corresponding to central extensions of $L_0$.
The structure constants 
can be regarded as fixed, numerical components of some complex tensors of the appropriate contravariant and covariant ranks. If needed,  
generic homogeneous basis elements are denoted $w_a$, $w_b$, $a,b=1,2,\cdots,m+n$; alternatively $w_i \equiv  x_i$ and $w_{n+r}\equiv y_r$, for indices in the specified ranges. Algebraic relations can then be written with the help of the usual notation for the grading, namely $|a|= \bar{0}$ for $a=1,2,\cdots,n$, and 
$|a|= \bar{1}$ for $a=n+1,n+2, \cdots,n+m$ (see for example \S \S \ref{subsec:PBWproof}, \ref{subsec:GeneralisedJacobi}). 
 \myeenv 

\noindent
Further important objects associated with $L$ and $U(I)$ are as follows. Let ${\mathfrak S}$ be the ideal generated by the symmetric and antisymmetric tensor products $L_{\bar{0}}\wedge L_{\bar{0}}$ and $L_{\bar{1}}\vee L_{\bar{1}}$ in $L\otimes L$ (that is, the spaces spanned by the elements $x_i\otimes x_j + x_j \otimes x_i$, $i \le j$, and 
$y_p\otimes y_q - y_q \otimes y_p$, $p < q$, for $i,j = 1,2,\cdots,n$ and $p,q = 1,2, \cdots, m$, respectively),
together with the subspace of $L_{\bar{0}}\otimes L_{\bar{1}} + L_{\bar{1}}\otimes L_{\bar{0}}$ spanned by 
$x_i\otimes y_p + y_p \otimes x_i$, for $i = 1,2,\cdots,n$ and $p = 1,2, \cdots, m$ ).
\mybenv{Definition} \textbf{Super symmetric tensor algebra:}\\
\label{def:SuperSymmAlg}\noindent
The super symmetric tensor algebra $S(L) = T(L)/{\mathfrak S}$ 
is the tensor algebra modulo these symmetric  and antisymmetric relations. 
\myeenv 
More generally, let $I_2 = I \cap L \otimes L$ be the projection of $I$ onto $L\otimes L$, and $J(I_2)$ be the corresponding two-sided ideal of $T(L)$ generated by $I_2$.
\mybenv{Definition} \textbf{Homogeneous quadratic Lie superalgebra:}\\
\label{def:HomoSuperAlg}\noindent
The homogeneous quadratic Lie superalgebra $U(I_2) = T(L)/J(I_2)$ 
is the tensor algebra modulo these homogeneous quadratic relations. 
\myeenv 

 To complete these introductory remarks, we refer to the notion of the Koszul complex associated with $U$, although we do not require this for the present work. Let $I^\perp \subset L^*\otimes L^* \cong (L\otimes L)^*$ be the quadratic annihilator of $I$, and $J({I}^\perp)$ the corresponding two-sided ideal. Define $U^! := T(L^*)/J({I}^\perp)$. Note that $(U^!)^*$ is a right $U^*$-module by the action $x\cdot \eta(\zeta):= x(\eta \zeta)$. Taking a basis $w_a$ for $L$, and the corresponding dual basis $w^a$ for $L^*$, it can be shown that the mapping $d\in End(U\otimes (U^!)^*)$ defined by
\[
d\cdot u\otimes v := \sum_a w_a u \otimes v w^a
\]
is a differential, that is, $d\circ d =0$ \cite{PolishchukPositselski2005qa}. 
\mybenv{Definition} \textbf{Koszul quadratic Lie superalgebra:}\\
\label{def:KoszulQuadSuperalg}\noindent
The quadratic Lie superalgebra $U$ is \emph{Koszul} iff the cohomology associated with the complex defined by the operator $d$  is trivial.
\myeenv
The role of the notion of Koszulness is shown in the following key theorem. Firstly note that the direct sum decomposition $I \subset L\otimes L + L + {\mathbb C}$ enables maps $\alpha:I_2 \rightarrow L$ and $\beta:I_2 \rightarrow {\mathbb C}$ to be defined such that 
\[
I = \{ x-\alpha(x) -\beta(x) | x \in I_2 \}.
\]
\mybenv{Theorem} \textbf{Poincar\'{e} Birkhoff Witt basis theorem (\cite{PolishchukPositselski2005qa}, Ch 5, Theorem 2.1):}\\
\label{thm:KoszulQuadSuperalg}\noindent
When $U(I_2)$ is Koszul, and the following conditions are satisfied:
\begin{align}
\mbox{(J1)}\qquad & \, \big( \alpha \otimes {\sf id} - {\sf id} \otimes \alpha \big) |_{(I_2 \otimes L \cap L \otimes I_2 )} \subset I_2; \nonumber \\
\mbox{(J2)}\qquad & \, \alpha \circ \big( \alpha \otimes {\sf id} - {\sf id} \otimes \alpha \big) |_{(I_2 \otimes L \cap L \otimes I_2 )} = 
- \big( \beta \otimes {\sf id} - {\sf id} \otimes \beta \big)|_{(I_2 \otimes L \cap L \otimes I_2 )}; \nonumber \\
\mbox{(J3)}\qquad & \, \beta \circ\big( \alpha \otimes {\sf id} - {\sf id} \otimes \alpha \big) |_{(I_2 \otimes L \cap L \otimes I_2 )} =0, \label{eq:GeneralisedJacobiAbstract}
\end{align}
we have the isomorphism 
\[
gr(U(I)) \cong U(I_2)
\]
\myeenv
The relations (J1--J3) are called the \emph{generalised Jacobi identities}.  As mentioned above, proving Koszulness requires an analysis of the structure of the tensor ideal $J(I)$, and as shown in \cite{PolishchukPositselski2005qa} requires certain distributivity conditions to hold on the subspaces (\ref{eq:GeneratedSubspaces}). Here instead we firstly (Lemma \ref{lem:KoszulQuadSuperalg} below) establish (J1), (J2) and (J3), and secondly (Lemma \ref{lem:PBWbasis}) prove that the homogeneous quadratic algebra $U(I_2)$ is a \emph{PBW algebra}, or of \emph{PBW type}, that is, it admits an ordered basis of monomials (a \emph{PBW basis}) in the generators $w_a$, under mild conditions. Now a result of Priddy (\cite{Priddy1970kr}, Theorem 5.3; see also \cite{Priddy:1970a}) states that a homogeneous quadratic algebra of {PBW type} is Koszul, and so the conditions for Theorem \ref{thm:KoszulQuadSuperalg} are established, and the isomorphism $gr(U(I)) \cong U(I_2)$ is guaranteed.
Therefore, the PBW basis also provides a basis for $gr(U(I))$, and hence, via the isomorphism of spaces, a basis also for $U$ itself -- the normal statement of the PBW basis theorem.   

\mybenv{Lemma} \textbf{Generalised Jacobi Identities for quadratic Lie superalgebras:}\\
\label{lem:KoszulQuadSuperalg}
\noindent
The bilinear relations (\ref{eq:GenJacobiComponents}) of a quadratic Lie superalgebra are equivalent to the generalised Jacobi identities (\ref{eq:GeneralisedJacobiAbstract}). \\
\textbf{Proof:} 
The first task is to identify generic elements of $I_2\otimes L\cap L\otimes I_2 $, by finding linear combinations of tensor products of basis elements from each space which match. 
In the standard Lie algebra case, the required combinations 
are easily found. Given basis elements $x_i$ and $x_j x_k- x_k x_j$  (dropping the tensor product symbol $\otimes$ for ease of writing), the cyclic sum can be written in two ways:
\begin{align} 
(x_ix_j\!-\!x_jx_i)x_k \!+\!(x_jx_k\!-\!x_kx_j)x_i \!+\!(x_kx_i\!-\!x_ix_k)x_j
= & \, x_i(x_j x_k \!-\! x_k x_j) \!+\! x_j(x_k x_i \!-\! x_i x_k) \!+\!x_k(x_i x_j \!-\! x_j x_i),\nonumber 
\end{align}
from which it is evident that the space spanned by such elements is simply the third rank exterior product. 
For Lie superalgebras, suitably graded cyclic sums with $y_py_q +y_qy_p$ are also included. 
For quadratic Lie superalgebras however, such elements of $I_2$ include additional terms (for example with $y_py_q +y_qy_p$ becoming $y_py_q +y_qy_p-d_{pq}{}^{k\ell}x_kx_\ell$ as above). Cyclic combinations of an arbitrary basis element tensored with such terms no longer produce an equality. However, in the quadratic Lie superalgebra case, it turns out that one can explicitly add and subtract terms on each side to compensate for these pairs. 

To go further we investigate (J1--J3) for a particular grading of the triple of basis elements from $\bar{1}\bar{1}\bar{1}$, $\bar{1}\bar{1}\bar{0}$, $\bar{1}\bar{0}\bar{0}$ and $\bar{0}\bar{0}\bar{0}$. Given (\ref{eq:QuadIdeals}) the maps
$\alpha$ and $\beta$ are (repeated indices summed)
\begin{align}
\alpha(x_i \otimes x_j - x_j \otimes x_i ) = & \,  {c_{ij}}^k x_k \equiv (\mbox{ad}_i x_j), \nonumber \\
\alpha(y_p \otimes y_q +y_q\otimes  y_p  -  {d_{pq}}^{k\ell} x_k\otimes  x_\ell) = & \, {b_{pq}}^k x_k\equiv {{\texttt b}_{pq}} , \nonumber \\
\beta(y_p \otimes y_q +y_q\otimes  y_p  -  {d_{pq}}^{k\ell} x_k\otimes  x_\ell) = & \, {a_{pq}}, \nonumber \\
\alpha(x_i\otimes  y_p - y_p \otimes x_i) = & \, \bar{c}_{ip}{}^q y_q \equiv (\mbox{ad}_i y_p). 
\label{eq:AlphaBetaMaps}
\end{align}
where the symbols $\mbox{ad}_i x_j$ and ${{\texttt b}_{pq}}$ are introduced to save space.
Let us consider for example $\bar{1}\bar{1}\bar{0}$; other cases can be handled similarly. 
The following identity 
\begin{align}
z_{pqi}  
= & \, (y_py_q+y_qy_p-{d}_{pq}{}^{k\ell}x_kx_\ell)x_i +(x_i y_q -y_q x_i)y_p + (x_i y_p -y_p x_i)y_q
-{d}_{pq}{}^{k\ell}(x_i x_k - x_k x_i)x_\ell \nonumber \\
= & \, x_i(y_py_q+y_qy_p-{d}_{pq}{}^{k\ell}x_kx_\ell) + y_p(y_qx_i-x_iy_q) +y_q(y_px_i-x_iy_p)
+ {d}_{pq}{}^{k\ell}x_k(x_i x_\ell - x_\ell x_i)
\label{eq:xyyElements}
\end{align}
shows that the $z_{pqi}$ span this grading of the intersection $L\otimes I_2 \cap I_2\otimes L $. For (J1) we consider
\begin{align}
(\alpha\circ {\sf id} - {\sf id}\otimes \alpha)(z_{pqi})=& \, 
{\texttt b}_{pq} x_i + (\mbox{ad}_i y_q) y_p + (\mbox{ad}_i y_p) y_q  -{d}_{pq}{}^{k\ell} (\mbox{ad}_i x_k)x_\ell -
\nonumber \\
& \, - x_i {\texttt b}_{pq}   +y_p (\mbox{ad}_i y_q) +y_q (\mbox{ad}_i y_p) -{d}_{pq}{}^{k\ell}x_k(\mbox{ad}_i x_\ell),
\label{eq:AlphaIdZ}
\end{align}
and rewrite this so as to expose the intersection with $I_2$:
\begin{align}
(\alpha\circ {\sf id} - {\sf id}\otimes \alpha)(z_{pqi})=& \,
 \big( (\mbox{ad}_i y_p) y_q \!+\! y_q (\mbox{ad}_i y_p) \!-\! d_{ip,q}{}^{k\ell} x_k x_\ell \big) +
\big( y_p (\mbox{ad}_i y_q) \!+\! (\mbox{ad}_i y_q) y_p \!-\! d_{p,iq}{}^{k\ell}x_k x_\ell \big)+ 
\nonumber \\
+ & \, \big( {\texttt b}_{pq} x_i \!-\! x_i {\texttt b}_{pq}\big) + \big( d_{ip,q}{}^{k\ell} x_k x_\ell \!+\! d_{p,iq}{}^{k\ell}x_k x_\ell  
\!-\! {d}_{pq}{}^{k\ell} (\mbox{ad}_i x_k)x_\ell \!-\! {d}_{pq}{}^{k\ell}x_k(\mbox{ad}_i x_\ell)
\big),
\nonumber
\end{align}
where new terms $d_{ip,q}{}^{k\ell}  :=
{\bar{c}_i}{}_p{}^r d_{rp}{}^{k\ell}$, $d_{p,iq}{}^{k\ell}  :=
{\bar{c}_i}{}_q{}^r d_{pr}{}^{k\ell}$ have been added and subtracted. Obviously the last term belongs to the symmetric product $L_0 \vee L_0$ and so must vanish if (J1) is to be fulfilled. It is readily checked that the component form of
\[
0= \big( d_{ip,q}{}^{k\ell} x_k x_\ell + d_{p,iq}{}^{k\ell}x_k x_\ell  
- {d}_{pq}{}^{k\ell} (\mbox{ad}_i x_k)x_\ell - {d}_{pq}{}^{k\ell}x_k(\mbox{ad}_i x_\ell),
\big)
\]
is nothing but the second equation of (\ref{eq:GenJacobiComponents}) above. Proceeding with
(J2), 
\begin{align}
\alpha \circ (\alpha\circ {\sf id} - {\sf id}\otimes \alpha)(z_{pqi})
= & \, (\mbox{ad}_i{\texttt b}_{pq}) - {\texttt b}_{ip,q} - {\texttt b}_{p,iq}
\end{align}
where ${\texttt b}_{ip,q}:={b}_{ip,q}{}^k x_k := \bar{c}_{ip}{}^r {b_{rq}}^k x_k$,
and ${\texttt b}_{p,iq}:={b}_{p,iq}{}^k x_k := \bar{c}_{iq}{}^r {b_{pr}}^k x_k$. Vanishing of this combination produces the component form corresponding to the third of equations (\ref{eq:GenJacobiComponents}).

The first and second equations of (\ref{eq:GenJacobiComponents}) are derived by considering (J1--J3) for the 
$\bar{0}\bar{0}\bar{0}$ and $\bar{0}\bar{0}\bar{1}$ sectors respectively, that is, rewriting combinations of $x_iyx_jx_k$ and 
$x_i x_j y_p$; the fourth and fifth entries are derived similarly by examining 
the $\bar{1}\bar{1}\bar{1}$ sector, that is, rewriting combinations of $y_py_qy_r$. 

\myeenv

The remaining requirement for the PBW basis theorem, the existence of a PBW basis, can be established under fairly general conditions on the partial ordering of ordered pairs of indices. The definition of the ordering, together with the proof of the following property in a slightly more general context is given in the Appendix, \S \ref{subsec:PBWproof} as Theorem \ref{pbw-algebra_thm}.
\mybenv{Lemma} \textbf{PBW basis for quadratic Lie superalgebras:}\\
\label{lem:PBWbasis}
\noindent
A homogeneous quadratic Lie superalgebra is of PBW type under any index ordering such that only those ${d_{pq}}^{k\ell}$ are nonvanishing for which $k,\ell$ precedes $p,q$.

\noindent
\textbf{Proof:} Using the text-book method of Serre \cite{Serre1992lie}. See appendix, \S 
\ref{subsec:PBWproof}. \myeenv
  
\section{Type I quadratic Lie superalgebras and Kac modules}
\label{sec:TypeIKacGould}
The class of quadratic algebras is much richer than that of Lie algebras. The same is true for the restricted class of quadratic Lie superalgebras that we study, and it is beyond the scope of this work to investigate their enumeration and classification. However, some remarks bearing on the structure can easily be derived in the notation that we have given, and this leads to a binary typology of quadratic Lie superalgebras analogous to that well known for Lie superalgebras \cite{Kac1977lsa}, as well as explicit methods for building the $d_{pq}{}^{k \ell}$. This is discussed in \S \ref{subsec:Classification} below, and in \S  \ref{subsec:KacModules} we take up the study of `type I' quadratic Lie superalgebras. This results in the analogue of the Kac module construction for superalgebras \cite{Kac1978rcls}, followed by a variant of this due to Gould \cite{Gould1989ar}, which gives an explicit construction of finite dimensional irreducible modules, both typical and atypical.
\subsection{Remarks on the structure of quadratic Lie superalgebras}
\label{subsec:Classification}
We assume from now on that the underlying Lie algebra $L_0$ is reductive.  For brevity we drop explicit reference to the tensor product $\otimes$ when working with $U(I)$ or its subalgebras. Given the significance of the generalised Jacobi identities established above, we also adopt the familiar
Lie algebra and Lie superalgebra notation for the structure relations in the form of brackets which have `nonlinear' parts on the right-hand side. Thus the second line of (\ref{eq:QuadIdeals}) now reads
\[
\{ y_p,y_q\} = {d_{pq}}^{k\ell}x_k x_\ell + {b_{pq}}^k x_k + a_{pq}
\]
where $\{\, ,\,\}$ is the anticommutator, $\{ u, v \} = uv + vu$ for odd homogeneous generators; for $L_0$ we have the usual commutator bracket ${[} x,x'{]} = xx'-x'x$. Now, because of the nonlinear brackets, there is no natural adjoint action $ad_w: L \rightarrow L$. However, (\ref{eq:QuadIdeals}) makes it clear that for $x\in L_0$ we can still define
 $ad_x:L_{\bar{1}} \rightarrow L_{\bar{1}}$ by 
\begin{align}
\label{eq:AdDefinition}
\mbox{ad}_x(y) = & \, xy-yx
\end{align}
for each $y\in L_1$, such that $\mbox{ad}_{[x,x']} = \mbox{ad}_x \mbox{ad}_{x'} - \mbox{ad}_{x'} \mbox{ad}_x$, turning $L_{\bar{1}}$ into a finite-dimensional $L_{\bar{0}}$-module.
A further simplification is to assume this is a representation equivalent to its contragradient.  
If the matrix representation of the even generators in the odd submodule is
$\pi(x_i)_p{}^q = -\bar{c}_{ip}{}^q$, by assumption there is a nondegenerate bilinear form $\Omega_{rs}$ with inverse $\Omega^{rs} $ such that 
\begin{align}
\label{eq:BalancedDefn}
\Omega^{qr}\pi(x_i)_r{}^s\Omega_{sp} = & \, - \pi(x_i)_p{}^q
\end{align}
which implements the equivalence between $\pi$ and its contragredient $-\pi^\top$. We adopt the following nomenclature:
 \mybenv{Definition} \textbf{Balanced quadratic Lie superalgebras}:
 \label{def:Balanced}\\ 
 A quadratic Lie superalgebra is called balanced if the odd submodule $L_{\bar{1}}$ is a real representation of the even Lie subalgebra $L_{\bar{0}}$. 
\myeenv

 \mybenv{Definition} \textbf{Typology of balanced quadratic Lie superalgebras}:
 \label{def:Typology}\\
Let $\lambda$ be the highest weight of an irreducible representation of $L_0$ with contragredient $\lambda^*$, and $V_0(\lambda)$, $V_0(\lambda^*)$ the corresponding $L_0$-modules. 
Balanced quadratic Lie superalgebras are categorised into the following types according to the odd submodule $L_{\bar{1}}$: \\[.3cm]
\hspace*{1cm}
\begin{tabular}[tbp]{rl}
\mbox{Type I$'$:}  & $L_{\bar{1}} = L_++L_- \cong V_0(\lambda) + V_0(\lambda^*)$, for $\lambda$ a complex representation, \\
& with $\{L_\pm,L_\pm\}=0$; \\
or \mbox{Type II:}  & $L_{\bar{1}} \cong V_0(\lambda)$, for $\lambda$ a real representation.\\
\end{tabular} \\
\mbox{}\myeenv

Note that the extra assumption $\{L_\pm,L_\pm\}=0$ for Type I$'$ is stronger than for Type I Lie superalgebras, where the classification theorems are enough to guarantee it. Here it is introduced in the absence of classification results to avoid pathologies. As a final comment on the general nature of quadratic Lie superalgebras, it is clear that the structure constants must be natural tensors of $L_0$ which bear a close relation to known invariants \cite{MacFarlaneMountain1998it}. 
Consider a Lie superalgebra or quadratic Lie superalgebra of balanced type with even and odd generators $x_i$, $y_p$ and structure constants $d_{pq}{}^{k\ell}$, $b_{pq}{}^{k}$ in the notation adopted above ($d \equiv 0$ in the Lie superalgebra case). Take quadratic and cubic invariants of $L_0$ to be of the form $C_2 = C^{ij}x_ix_j$, $C_3 = C^{ijk}x_ix_jx_k$. 
The following  result is easily proven and provides a guaranteed explicit construction of $d_{pq}{}^{k\ell}$, as an extension of the (known) result for Lie superalgebras, which is here recalled:
\mybenv{Lemma} \textbf{Structure constants for Lie superalgebras and quadratic Lie superalgebras}:
\label{lem:StructureConsts}\\
The following forms for the structure constants fulfil the Jacobi and generalised Jacobi identities for Lie superalgebras and quadratic Lie superalgebras:
\begin{enumerate}
\item \hspace*{1cm}$b_{pq}{}^i = C^{ij}\pi(x_j)_p{}^r \Omega_{rq}$;
\item \hspace*{1cm}$d_{pq}{}^{k \ell} = C^{mk\ell}\pi(x_m)_p{}^r \Omega_{rq}$;
\end{enumerate}
\textbf{Proof}: The required Jacobi identities are tantamount to proving the invariance of the given Casimir operators. 
For the superalgebra case we use the Casimir invariant $C_2$ and the following universal object 
\cite{OBrienCantCarey1977} in $U_0 \otimes U_0$, the cut coproduct
\begin{align}
\label{eq:CutCoproduct}
\Delta C'_2 = & \, \textstyle{\frac 12}(\Delta C_2 - C_2 \otimes {\sf id} - {\sf id} \otimes C_2) \equiv C^{ij}x_i\otimes x_j
\end{align}
where $\Delta x = x\otimes {\sf id} + {\sf id} \otimes x$ is the usual coproduct Lie algebra homomorphism.
The object $\pi \otimes {\sf id} \circ \Delta C'_2$ commutes with $\pi \otimes {\sf id} \circ \Delta x =
\pi(x)\otimes {\sf id} + {\sf id}\otimes x$ in $End(L_{\bar{1}})\otimes U_0$; note that the form
$b_{pq}{}^i$ is the matrix element of the bilinear form $\Omega$ composed with this map, namely
$C^{ij}\pi(x_j)_p{}^r \Omega_{rq}$:
\[
C^{ij}{[}\pi(x_k),\pi(x_i){]} \otimes x_j + C^{ij}\pi(x_i) {[}x_k,x_j{]} =0
\]
or taking matrix elements
\[
C^{ij} \pi(x_k)_p{}^t \pi(x_i)_t{}^r\Omega_{rq}x_j - C^{ij} \pi(x_i)_p{}^t \pi(x_k)_t{}^r \Omega_{rq} x_j
+ C^{ij} \pi(x_i)_p{}^r \Omega_{rq} c_{kj}^\ell x_\ell =0.
\]
We identify $b_{pq}{}^i$ as above, and also $\pi(x_k)_p{}^t = -\bar{c}_{kp}{}^t$. For the second term we need the intertwining property of $\Omega_{pq}$, namely
\[
\Omega^{pr} \pi(x_k)_r{}^s \Omega_{sq} = -\pi(x_k)_q{}^p,\qquad 
\mbox{or} \qquad \pi(x_k)_t{}^r\Omega_{rq} = - \pi(x_k)_q{}^r \Omega_{tr}.
\]
The invariance condition thus becomes
\[
-\bar{c}_{kp}{}^t b_{tq}{}^j x_j -\bar{c}_{kq}{}^r b_{pr}{}^j x_j  + b_{pq}{}^\ell c_{k\ell}{}^j x_j = 0
\]
which is just the second line of (\ref{eq:GenJacobiComponents}) above.

In the quadratic Lie superalgebra case, the same procedure works starting with the invariance of the cut coproduct 
\[
\Delta C'_3 = \textstyle{\frac 13}(\Delta C_3 - C_3 \otimes {\sf id} - {\sf id} \otimes C_3)
\]
of a cubic Casimir $C_3 = C^{k\ell m}x_k x_\ell x_m$, namely (taking account of the symmetry of $C^{ijk}$ and the mixed nature of the terms arising from the coproduct after removing $C_3 \otimes {\sf id}$ and ${\sf id}\otimes C_3$),
\[
{[}\pi(x_i)\otimes {\sf id} + {\sf id}\otimes x_i , C^{k\ell m}\pi(x_k) \otimes x_\ell x_m + 
C^{k\ell m}\pi(x_k x_\ell) \otimes x_m  {]} = 0.
\]
The new feature arising is that this expression in $End(L_{\bar{1}}) \otimes U_0$ is graded by degree in $U_0$, and invariance requires both linear and quadratic parts to vanish separately. Taking matrix elements in the projection onto the quadratic part,
\[
{[}\pi(x_i)\otimes {\sf id} + {\sf id}\otimes x_i , C^{k\ell m}\pi(x_k) \otimes x_\ell x_m {]} = 0
\]
and using the intertwining property of the map $\Omega$ as above, yields the first line of (\ref{eq:GenJacobiComponents}).
\myeenv

For a more complete analysis along these lines, further considerations on a case-by-case basis, for example whether $C_3$ is given in irreducible form, would be needed. The object (\ref{eq:CutCoproduct}) will play a crucial role in the projection operator techniques to be used in \S \ref{subsec:PolynomialIdentities}.
\subsection{Enveloping algebra for type I quadratic Lie superalgebras and Kac modules}
\label{subsec:KacModules}
The definition of type I$'$ quadratic Lie superalgebras means that $L_{\overline{1}} \cong
V_0(\lambda) + V_0(\lambda^*)$ for some complex irreducible $L_0$-module with highest weight $\lambda$. The condition
$\{ L_\pm, L_\pm \} = 0$ is consistent with a 
${\mathbb Z}$-grading of $L$ such that $L_{+1}=L_+ \cong V_0(\lambda)$, $L_{+1}=L_- \cong V_0(\lambda^*)$, so that 
$L \cong L_{\overline{0}} + L_{\overline{1}}\cong L_0 + L_+ + L_-$. Define $U_0 = U(L_0)$, $U_\pm = U(L_\pm)$ as subalgebras of $U$. With respect to a basis $\{ \overline{Q}{}^1, \overline{Q}{}^2, \cdots, \overline{Q}{}^d, \}$ for $L_+$, and a dual basis $\{ {Q}{}_1, {Q}{}_2, \cdots, {Q}{}_d, \}$ for $L_-$, note that
\begin{align}
U_+ \cong & \, \left<\,  \overline{Q}{}^{i_{1}} \overline{Q}{}^{i_{2}} \cdots \overline{Q}{}^{i_{k}}, \, 
i_1\! <\!i_2\! <\!\cdots\! <i_k,\, 1\!\le\! k\! \le\! d \, \right>,  \quad \mbox{and} \quad \nonumber \\
U_- \cong & \, \left<\,  {Q}_{i_{1}} {Q}_{i_{2}} \cdots {Q}_{i_{k}}, \, i_1\! <\!i_2\!<\!\cdots\! <i_k,\, 1\!\le\! k\! \le\! d \, \right>,
\label{eq:UpUmGenerators}
\end{align}
with each of dimension $2^d$ if $\mbox{dim}(V_0(\lambda))=d$. We note also the canonical elements
\begin{align}
{S} := & \, 
 {Q}_{1} {Q}_{2} \cdots {Q}_{d}, \qquad  \overline{S} := 
 \overline{Q}{}^{1} \overline{Q}{}^{2} \cdots \overline{Q}{}^{d},
\end{align}
which are, in particular, one dimensional $L_0$-modules under the natural extension of the adjoint action (compare (\ref{eq:AdDefinition})). Let the respective (highest) weights be $\mp 2\rho_1$, that is, we take
${[}h, \overline{S}{]} = 2\rho_1(h)= -{[}h, S{]}$ for $h$ belonging to a Cartan subalgebra of $L_0$.
From the PBW basis theorem it follows that
\begin{align}
U \cong &\, U_-U_0U_+ \equiv U_-\overline{U}_+, \quad \mbox{and} \nonumber \\
U \cong &\, U_+U_0U_- \equiv U_+ \overline{U}_-. \nonumber 
\end{align}
This is easily checked by considering basis elements of $U_0$, $U_\pm$ and re-ordering the appropriate products into the standard ordered monomial basis of $U$, or \emph{vice versa} by showing that the latter monomials can be expressed as combinations of the re-ordered product forms.

We introduce finite-dimensional induced representations 
as follows \cite{Kac1977lsa,Kac1978rcls}. Let $V_0(\Lambda)$ be a finite-dimensional $L_0$-module with (dominant integral) highest weight $\Lambda$. Make $V_0(\Lambda)$ into a $\overline{U}_+$-module by declaring $y v =0$ for $y \in L_+$, for all states $v \in V_0(\Lambda)$. Define
\[
\overline{V}(\Lambda) := U \otimes_{\overline{U}_+} V_0(\Lambda), \quad  
\overline{V}(\Lambda)  \cong U_- \otimes V_0(\Lambda).
\]
\mybenv{Definition} \textbf{Kac modules for quadratic Lie superalgebras}:\\
\label{def:KacModules}
Let ${\mathfrak I}$ be the maximal invariant submodule of  $\overline{V}(\Lambda)$. The Kac module ${V}(\Lambda)$ is the irreducible factor module defined as ${V}(\Lambda):=\overline{V}(\Lambda)/{\mathfrak I}$.
\mbox{} \myeenv
A crucial point in relation to this construction is due to Gould \cite{Gould1989ar}: namely, that the (necessarily irreducible) module cyclically generated by the states
$S \otimes v$, $v \in V_0(\Lambda)$, is a factor module of \emph{every} invariant submodule of $\overline{V}(\Lambda)$ . By appropriate reorganisation of weight labels, this observation leads to the following direct construction of $V(\Lambda)$ itself:
\mybenv{Theorem} \textbf{Modified Kac module construction by levels}:\\
\label{thm:ModifiedKac}
Let $V_0(\Lambda - 2\rho_1)$ be a lowest weight $L_0$-module, regarded as a $\overline{U}_-$-module by declaring $y v =0$ for $y \in L_-$, for all states $v \in V_0(\Lambda-2\rho_1)$. Then the Kac module $V(\Lambda)$ is isomorphic to 
the induced module $U\overline{S}\otimes_{\overline{U}_-}V_0(\Lambda-2\rho_1)$, that is, we have
\begin{align}
V(\Lambda) \cong & \, {U}_- \cdot \overline{S} \otimes_{\bar{U}_-} V_0({\Lambda-2\rho_1}), \quad \mbox{or} \nonumber \\
V(\Lambda) \cong & \, \langle Q_{i_1} Q_{i_2} \cdots Q_{i_k} \overline{S} \! \otimes \!v\rangle, \, 
i_1 <i_2 <\cdots <i_k,\, 0\le k \le n, v\in V_0({\Lambda-2\rho_1}). 
\label{eq:GeneralForm}
\end{align}
\textbf{Proof:} See Gould \cite{Gould1989ar}.
\mbox{} \hfill \myeenv
Each of the above subspaces of $V(\Lambda)$ for $k$ fixed is referred to as the subspace at \emph{level} $k$, $0 \le k \le n$. By analogy with the Lie superalgebra case, the module $V(\Lambda)$ is said to be \emph{typical} if the level $k=n$ subspace is nontrivial, $\langle S \overline{S}\otimes_{\overline{U}_-}v, v\in V_0({\Lambda-2\rho_1})\rangle  \ne \phi$; otherwise $V(\Lambda)$ is said to be \emph{atypical}.  


\section{The quadratic Lie superalgebra $gl_2(n/1)$ and atypicality conditions}
\label{sec:Atypicals}
In this section we wish to apply the results of  \S\S \ref{sec:FormalPBW},\ref{sec:TypeIKacGould}, and especially \S \ref{subsec:KacModules}, Theorem \ref{thm:ModifiedKac},  to a specific quadratic Lie superalgebra, as a concrete case study. Several instances of (quadratic versions of) `polynomial super $gl(n)$' algebras were given in \cite{JarvisRudolph2003} using the notation $gl_2(n/\lambda + \lambda^*)$, to indicate that the even subalgebra is $gl(n)$, and the odd submodule 
$L_{\overline{1}} \simeq V_0(\lambda) + V_0(\lambda^*)$ has the type I structure as in definition \ref{def:Typology} above. Specific examples of $\lambda$ given in \cite{JarvisRudolph2003} are 
$\Lambda_1$, $\Lambda_3$, or $3\Lambda_1$, where the odd generators are either in the fundamental defining representation, or antisymmetric or symmetric rank 3 tensor representations of $gl(n)$, respectively (denoted ${\{} 1{\}}$, ${\{} \!1,\!1,\!1{\}}$ and ${\{} 3{\}}$, in partition form), together with their contragredients. The last two cases will be discussed briefly in the concluding remarks, \S \ref{sec:Examples}.
Here we take up the case of odd generators in the fundamental representation, which we denote here simply $gl_2(n/1)$, by analogy with the Lie superalgebra case. 


\subsection{General results and structure for $gl_2(n/1)$}
\label{subsec:GeneralResults}
The even generators are the generators of $gl(n)$ in Gel'fand form, $E^i{}_j$, $i,j = 1,2,\cdots, n$, and the odd generators are denoted $Q_i$, $\overline{Q}{}^j$, $i,j = 1,2,\cdots,n$. In the notation of \S \ref{subsec:KacModules}, we have $gl_2(n/1) \cong L\cong L_- \!+\! L_0\! +\! L_+$ with $L_0 \cong L_{\bar{0}} \cong gl(n)$, $L_{\bar{1}} \cong L_+ \!+\! L_-$ and $L_-$, $L_+$ spanned by
$Q_i$, $\overline{Q}{}^j$ respectively. The $E^i{}_j$ have the standard commutation $gl(n)$ relations 
\begin{align}
\label{eq:GLnRelations}
{[}E^i{}_j, E^k{}_\ell {]} = & \, \delta_j{}^k E^i{}_\ell - \delta^i{}_\ell E^k{}_j, 
\end{align}
while the components of $\overline{Q}$ form a vector, and the $Q$ a contragredient vector, under the adjoint action of $gl(n)$,
\begin{align}
\label{eq:VectorOpRelations}
{[}E^i{}_j, \overline{Q}{}^k{]} = & \, \delta^k{}_j \overline{Q}{}^i , \qquad  {[}E^i{}_j, Q_k{]} = -\delta^i{}_kQ_j, 
\end{align}
with ${\{}Q_i,Q_j{\}} = 0 = {\{} \overline{Q}{}^i,\overline{Q}{}^j{\}}$. Finally the quadratic bracket relations between $Q$ and $\overline{Q}$ read
\begin{align}
{\{} \overline{Q}{}^i, Q_j{\}} = & \, (E^2)^i{}_j - \langle E \rangle E^i{}_j -\textstyle{\frac 12} \delta^i{}_j 
\big( \langle E^2\rangle - \langle E\rangle^2 +(n\!-\!1) \langle E \rangle \big) + \texttt{c} \delta^i{}_j, 
\label{eq:gln1StructConsts}
\end{align}
where $(E^2)^i{}_j =
E^i{}_kE^k{}_j$, $\langle E \rangle := \sum_i E^i{}_i$, $\langle E^2\rangle = \sum_i  (E^2)^i{}_i$ are the linear and quadratic Casimirs, and $\texttt{c}$ is a central term. Note that for the linear Casimir we have ${[}\langle E \rangle, \overline{Q}{}^i{]}= \overline{Q}{}^i$, and 
${[}\langle E \rangle, Q_j{]}=-Q_j$. In \cite{JarvisRudolph2003} it has been established that the structure constants implicit in these relations (\ref{eq:gln1StructConsts}) obey (\ref{eq:GenJacobiComponents}). As far as the factorisation of the enveloping algebra is concerned (see \S \ref{subsec:KacModules}) (\ref{eq:UpUmGenerators}) applies directly so that each of $U_\pm$ is finite-dimensional with $2^n$ components.

The close relationship between $gl_2(n/1)$ and the Lie superalgebra $sl(n/1) \cong A(n\!-\!1,0)$ is illustrated by the fact that the $n=2$ quadratic case is degenerate:
\mybenv{Lemma} \textbf{Structure of $gl_2(2/1)$}:\\
The quadratic Lie superalgebra $gl_2(n/1)$ for $n=2$ is degenerate and isomorphic to the Lie superalgebra $sl(2/1)\cong A(1,0)$.\\
\textbf{Proof}:\\
By direct evaluation of (\ref{eq:gln1StructConsts}) for the case $n=2$, we find that the contributions to $d_{p,q}{}^{k,\ell}$ vanish:
\begin{align}
{\{}\overline{Q}{}^1, Q_1{\}} = & \, -E^2{}_2 + \texttt{c}, \qquad  {\{}\overline{Q}{}^1, Q_2{\}} =  E^1{}_2, \qquad \qquad \nonumber \\
{\{}\overline{Q}{}^2, Q_1{\}} = & \,E^2{}_1,  \qquad \qquad  {\{}\overline{Q}{}^2, Q_2{\}} = -E^1{}_1 + \texttt{c}.\hfill \nonumber 
\end{align}
On the other hand the Lie superalgbra $sl(2/1)$ has dimension $8$, with odd generators $R_a$, $\overline{R}^b$, $a,b=1,2$ transforming as doublet and conjugate doublet under the even part $sl(2) + u(1)$ generated by ${\mathbf J}= (J_1,J_2,J_3)$, $Z$, with standard (anti)commutation relations
\begin{align}
{[} {\mathbf J}, R_a {]} = & \, -\textstyle{\frac 12}\mbox{\boldmath{$\sigma$}}_a{}^bR_b, \quad 
{[} Z, R_a {]} =  - R_a ; \nonumber \\
{[} {\mathbf J}, \overline{R}^a {]} = & \, +\textstyle{\frac 12}\overline{R}^b\mbox{\boldmath{$\sigma$}}_b{}^a,
\quad {[} Z, \overline{R}^a {]} =  +\overline{R}^a; \nonumber \\
{[} J_i, J_j{]} = & \, \varepsilon_{ijk}J_k, \qquad \quad {\{} R_a, \overline{R}^b{\}} = ({\mathbf J}\!\cdot\!\mbox{\boldmath{$\sigma$}})_a{}^b +\delta_a{}^b Z, \nonumber
\end{align} 
where $i,j,k = 1,2,3$, and with $\varepsilon_{ijk}$ the usual Levi-Civita symbol (the sign of the permutation $(ijk)$). These provide the structure constants $c_i{}^{jk}$ of $sl(2)$, while the remaining structure constants ($\overline{c}_{ip}{}^q$ and $b_{pq}{}^k$ in the notation of (\ref{eq:GenJacobiComponents})) given in terms of standard Pauli matrices\footnote{
$\mbox{\boldmath{$\sigma$}}=(\sigma_1,\sigma_2,\sigma_3)$ with
$\sigma_1 = \left(\begin{array}{cc}0&1\\1&0\end{array}\right)$,
$\sigma_2 = \left(\begin{array}{cc}0&\!\!-\!i\\i&0\end{array}\right)$,
$\sigma_3 = \left(\begin{array}{cc}1&0\\0&\!\!-\!1\end{array}\right)$. 
}. The identification is given via  $Q_a \rightarrow R_a$, $\overline{Q}^a \rightarrow \overline{R}^a$,
$E^1{}_2 \rightarrow J_+$, $E^2{}_1 \rightarrow J_-$, with $\textstyle{\frac 12}(E^1{}_1-E^2{}_2) \rightarrow J_3$,
$-\textstyle{\frac 12}(E^1{}_1+E^2{}_2) + \texttt{c} \rightarrow Z$.
\\ \mbox{} \myeenv

Roots and weights of $gl(n)$ are introduced as follows. Take the standard Cartan algebra $E^1{}_1, E^2{}_2, \cdots, E^n{}_n$ and positive simple root vectors $E^1{}_2$, $E^2{}_3$, $\cdots$, $E^{n\!-\!1}{}_n$. Consider Euclidean vectors in ${\mathbb  C}^n$ with basis ${e_i } $ and standard inner product $(e_i,e_j)=\delta_{ij}$. 
The trace form in the defining representation associates each root vector $E^{i}{}_j$ ($i \ne j$) with $e_i-e_j$. A $gl(n)$-module $V_0(\lambda)$ is a highest weight representation if there is a vector $v_\lambda$ with weight
$\lambda_1 e_1 + \lambda_2 e_2 + \cdots +\lambda_n e_n \equiv (\lambda_1,\lambda_2,\cdots, \lambda_n)$ annihilated by $E^{i}_j$, $i<j$ with $\lambda_1 \ge \lambda_2 \ge \cdots \ge \lambda_n$. $V_0(\lambda)$ is finite dimensional if $\lambda$ is dominant integral, with $\lambda_i - \lambda_{i\!+\!1} \in {\mathbb Z}^+$ for $i=1,2,\cdots, n\!-\!1$. In this notation the half-sum of positive roots is
$\rho_0 = \textstyle{\frac 12}\big( (n\!-\!1), (n\!-\!3), \cdots, (n\!+\!1-2r), \cdots, -(n\!-\!1)\big)$. Weights of the vector $\overline{Q}$ are evidently those of the fundamental representation, simply $(1,0,\cdots, 0)$, $(0,1,0,\cdots)$,
$\cdots$, $(0,0,\cdots,0,1)$ and the weights of the components $Q$ are the negative of these. The quantity $\rho_1$, the half-sum of positive odd weights, is $\rho_1 = \textstyle{\frac 12}\big(1,1,\cdots, 1,1)$ (which coincides with the usual half-sum of positive odd roots in the $gl(n/1)$ case). Finally note that the eigenvalues of the quadratic and linear Casimirs in the highest weight representation $V_0(\lambda)$ are $C_1 = \sum_r \lambda_r$, and $C_2 = \sum_r \lambda_r(\lambda_r +n+1-2r)$, respectively.

We now proceed with the concrete steps elaborated in \S \ref{subsec:KacModules} for the construction of irreducible modules of $gl_2(n/1)$. Let $\Lambda$ be a dominant integral $gl(n)$-weight and consider the corresponding weight $\Lambda' := \Lambda-2\rho_1$ (which is also dominant integral). Take a lowest-weight $gl(n)$ module $V_0(\Lambda')$. Introduce as before $\overline{S}$ in the form
\begin{align}
 \overline{S} = & \,
{ \frac{1}{n!}}  
\varepsilon_{i_1 i_2 \cdots i_n} \overline{Q}^{i_1} \overline{Q}^{i_2} \cdots \overline{Q}^{i_n}, \nonumber 
\end{align}
with the help of the Levi-Civita tensor $\varepsilon_{i_1 i_2 \cdots i_n}$.
Then from \S \ref{subsec:KacModules}, Theorem \ref{thm:ModifiedKac} and following 
 \cite{Gould1989ar,GouldBrackenHughes1989br}, we have for the irreducible $gl_2(n/1)$ Kac module associated with $\Lambda$, 
\begin{align}
\label{eq:ExplicitForm}
 V(\Lambda) \cong & \, \left< Q_{i_1} Q_{i_2} \cdots Q_{i_k} \overline{S} \! \otimes \!v, \, 
i_1 <i_2 <\cdots <i_k,\, 1\le k \le n, v\in V_0({\Lambda'}) \right>,
\end{align}
with the tensor product of type $\otimes_{\overline{U}_-}$. This equation (\ref{eq:ExplicitForm}) is the key construct from which the structure of the irreducible module can be derived. As we shall see, tensor projection methods can be applied, which explicitly give rise to patterns of vanishing of certain combinations of operators 
$Q_{i_1} Q_{i_2} \cdots Q_{i_k}$ applied to $\overline{S}\otimes v$, depending on the components of the highest weight vector $\Lambda = (\Lambda_1, \Lambda_2, \cdots, \Lambda_n)$, and central charge $\texttt{c}$. Firstly we record important basic relations between the generating multinomials of (\ref{eq:UpUmGenerators}) above. In addition to $\overline{S}$ above, introduce the following auxiliary quantities:
\begin{align}
\label{eq:Auxiliaries}
 \overline{S} = & \,
{ \frac{1}{n!}}  
\varepsilon_{i_1 i_2 \cdots i_n} \overline{Q}^{i_1} \overline{Q}^{i_2} \cdots \overline{Q}^{i_n}, \nonumber \\
\overline{S}_{i_1 i_2 \cdots i_k} :=& \, 
{ \frac{(-1)^{\frac 12k(k\!-\!1)}}{(n\!-\!k)!} } 
                                                          \varepsilon_{i_1 i_2 \cdots i_k i_{k+1} \cdots i_n}\overline{Q}^{i_{k+1}} \overline{Q}^{i_{k+2}} \cdots \overline{Q}^{i_n}, \nonumber \\
                  \overline{Q}^{i_1} \overline{Q}^{i_2} \cdots \overline{Q}^{i_k} = & \, (-1)^{\frac 12(n\!-\!k)(n\!+\!k\!-\!1)}  \varepsilon^{i_1 i_2 \cdots i_k i_{k+1} \cdots i_n}
                  \overline{S}_{{i_{k+1}}  {i_{k+2}}  \cdots {i_{n}} }.
\end{align}
\mybenv{Lemma} \textbf{Calculus of multinomial odd generators}:\\
The quantities (\ref{eq:Auxiliaries}) satisfy the following relations:
\begin{align}
\overline{Q}{}^i \overline{S} = & \, 0 , \nonumber \\
\overline{Q}{}^i\overline{S}_j =  & \, {\delta^i}_j \overline{S},\nonumber \\ 
\overline{Q}{}^i \overline{S}_{jk} = & \, {\delta^i}_j \overline{S}_k - {\delta^i}_k \overline{S}_j, \nonumber \\
\overline{Q}{}^i \overline{S}_{jk \ell \cdots } = & \,  {\delta^i}_j \overline{S}_{k \ell \cdots }  - {\delta^i}_k \overline{S}_{j \ell \cdots } 
                                                                                  +  {\delta^i}_\ell \overline{S}_{jk  \cdots } + \cdots, \nonumber \\
{[}{E^i}_j, \overline{S}_{i_1 i_2 \cdots i_k} {]} = & \, {\delta^i}_j\overline{S}_{i_1 i_2 \cdots i_k} - 
                                                                          {\delta^i}_{i_1} \overline{S}_{j i_2 \cdots i_k} - \cdots, 
\label{eq:QbarSbarProps}
\end{align}
whence ${[}\langle E \rangle, \overline{S}_{i_1 i_2 \cdots i_k} {]} = (n\!-\!k)  \overline{S}_{i_1 i_2 \cdots i_k}$. Moreover, we have
\begin{align}
{[}Q_i, \overline{S}{]} = & \, \overline{S}{}_k A^k{}_i \equiv \overline{A}_i{}^k \overline{S}_k, \nonumber \\
{[}Q_i, \overline{S}_j {]} = & \,
 \overline{S}{}_{k\ell} B^{k\ell}{}_{ij}, 
\label{eq:QSbarCommutators}
\end{align}
where the quantities $A$, $B$ are
\begin{align}
 A^i{}_j = & \, (E^2)^i{}_j - (\langle E\rangle \!+\!(n\!-\!2)) E^i{}_j -\textstyle{\frac 12} \delta^i{}_j 
\big( \langle E^2\rangle - \langle E\rangle^2 -(n\!-\!3) \langle E \rangle \big) + (c-(n\!-\!1)) \delta^i{}_j,
\nonumber \\
{B}^{k \ell}{}_{ij} = & \,
\big((E^2)^k{}_i \delta^\ell{}_j -(k\ell) \big) - \big( E^k{}_i \delta^\ell{}_j -(k\ell) \big)(\langle E\rangle+\!n\!-\!3)
+ \big(\delta^k{}_i E^\ell{}_j -(k\ell) \big) \nonumber \\
& \, 
-\textstyle{\frac 12}\big(\delta^k{}_i \delta^\ell{}_j -(k\ell) \big)(\langle E^2\rangle-\langle E\rangle^2-(n\!-\!5)\langle E\rangle) + \big(\delta^k{}_i \delta^\ell{}_j -(k\ell) \big)(c\!-\!(n\!-\!2)). 
\label{eq:AsAndBs}
\end{align}
\textbf{Proof}:\\
(\ref{eq:QbarSbarProps}) is proven
by explicit calculation using the anticommutation and commutation relations. For the first step of (\ref{eq:QSbarCommutators}), (\ref{eq:AsAndBs}) we have from 
$\overline{Q}{}^m\overline{S}=0$, the bracket
\begin{align}
0 = & \, {[}Q_i, \overline{Q}{}^m\overline{S}{]} = {\{}Q_i, \overline{Q}{}^m{\}} \overline{S} - 
          \overline{Q}{}^m {[}Q_i, \overline{S}{]} \nonumber \\
\mbox{thus} \qquad \qquad 
{\{}Q_i, \overline{Q}{}^m{\}} \overline{S} = & \, \overline{Q}{}^m \overline{S}{}_k A^k{}_i
 \equiv \overline{S} A^m{}_i;
 \nonumber  \\
\mbox{so} \qquad  \qquad \qquad 
\overline{S} A^m{}_i = & \,  \big( (E^2)^m{}_i - \langle E \rangle E^m{}_i -\textstyle{\frac 12} \delta^m{}_i 
\big( \langle E^2\rangle - \langle E\rangle^2 +(n\!-\!1) \langle E \rangle \big) + c \delta^m{}_i \big) \overline{S}
\nonumber
\end{align}
from which the form of $A^m{}_i$, and subsequently $\overline{A}_i{}^m$, can be read off by appropriate shifting, for example $E^k{}_\ell \overline{S} = \overline{S}\big( E^k{}_\ell + n \delta^k{}_\ell \big)$, and similarly
$\overline{S}{}_k E^k{}_i = \big( (n-1)\delta_i{}^k -\overline{E}_i{}^k\big)\overline{S}{}_k$.
The second step proceeds analogously, starting firstly with ${[} Q_i, \overline{S}{}_j{]} = \overline{S}{}_{k\ell}
B^{k \ell}{}_{ij}$ and hence $\overline{Q}{}^m {[} Q_i, \overline{S}{}_j{]} = 2 \overline{S}_\ell B^{m \ell}{}_{ij}$; on the other hand noting 
\begin{align}
{[}Q_i, \overline{Q}{}^m \overline{S}{}_j {]} = & \, {\{}Q_i, \overline{Q}{}^m{\}} \overline{S}{}_j  - \overline{Q}{}^m {[}Q_i,  \overline{S}{}_j {]} \nonumber
\end{align}
we can rearrange using also $\overline{Q}{}^m \overline{S}{}_j = \delta^m{}_j \overline{S}$ to get
\begin{align}
2 \overline{S}_\ell B^{m \ell}{}_{ij} = & \, 
\big( (E^2)^m{}_i - \langle E \rangle E^m{}_i -\textstyle{\frac 12} \delta^m{}_i 
\big( \langle E^2\rangle - \langle E\rangle^2 +(n\!-\!1) \langle E \rangle \big) + c \delta^m{}_i \big)\overline{S}{}_j - \delta^m{}_j \overline{S}_k A^k{}_i
\nonumber
\end{align}
which allows $B^{m \ell}{}_{ij}$ to be inferred by shifting the terms in even generators to the right hand side as before.
\\
\mbox{} \myeenv

\subsection{Review of polynomial characteristic identities and shift operator analysis for $gl(n)$}
\label{subsec:PolynomialIdentities}

In \S \ref{subsec:Classification}, Lemma \ref{lem:StructureConsts}, (\ref{eq:CutCoproduct}), we gave following
\cite{OBrienCantCarey1977} a very general way of identifying objects in the enveloping algebra $U(gl(n))$
when acting on an arbitrary $gl(n)$-module $V_0(\lambda)$.
In the case of the standard $gl(n)$ Gel'fand generators, the $-E^i{}_j$ are the matrix elements of the object (\ref{eq:CutCoproduct}) in $End(V_0(\Lambda_1^*)\otimes End(V_0(\lambda)))$. As has been shown in 
\cite{green1971ci,OBrienCantCarey1977} (see also \cite{Gould:1984} and \cite{Gould1989ar,GouldBrackenHughes1989br} for applications to Lie superalgebras), these objects satisfy a certain polynomial characteristic identity, of the form $\prod_s (E-\alpha_s) =0$, where general matrix powers of the array $E$ are defined (extending the notation used already) by $(E^0){}^i{}_j = \delta^i{}_j$, $(E^{m\!+\!1}){}^i{}_j = E^i{}_k (E^{m}){}^k{}_j$.
The roots are dependent on the components of the highest weight $\lambda$ via $\alpha_s = \lambda_s+n-s$, $s=1,2,\cdots, n$. Dually there is an analogue identity $\prod_s (\overline{E}-\overline{\alpha}_s) =0$ for the array $\overline{E}_i{}^j =
- E^j{}_i$ with its matrix powers defined analogously, and $\overline{\alpha}_s = n - 1 - \lambda_s$.

Let $V$, $W$ be $gl(n)$-modules, and consider a \emph{tensor operator} $T$ -- that is, a map $T: V \rightarrow End(W)$ which intertwines the action of the Lie algebra, in the sense that ${[} x, T(v){]} = -T(x\cdot v)$ for each $v\in V$. 
Since $x \cdot T(v)w = T(-x\cdot v)w + T(v)x\cdot w$, the space $T(V)W$ is isomorphic to the tensor product $V^*\otimes W$. 
If $V= {\mathbb C}^n$, identified as the fundamental representation $\Lambda_1$, the object $T$ is termed a \emph{vector operator}, and the $T_i := T(e_i)$ with respect to the standard basis form the \emph{components} of $T$.
Correspondingly if $V={\mathbb C}^n$ is identified with the dual $\Lambda_1^*$ rather than the defining representation $\Lambda_1$ of $gl(n)$, the corresponding object, say $\overline{T}$, is called a \emph{contravariant} vector operator, and its components are $\overline{T}{}^i := \overline{T}{}(f^i)$ with respect to the standard dual basis. 

For (contravariant) vector operators, the polynomial characteristic identity can be used to provide a resolution into a sum of shift operators, themselves vector operators, using projection operators, 
\begin{align}
T_i = & \,  \sum_{s=0}^nT_i[s], \qquad  T_i[s]:= \overline{P}[s]_i{}^k T_k = T_k {P}[s]^k{}_i,
\label{eq:ShiftDefinitions}
\end{align}
with similar expressions for $\overline{T}{}^i $, where each projection operator is built from the appropriate monomial factors of the polynomial identity, namely
\[
{P}[s] = \frac{\prod_{s'\ne s} (E-\alpha_{s'})}{\prod_{s' \ne s} (\alpha_s-\alpha_{s'})},
\quad
\overline{P}[s] = \frac{\prod_{s'\ne s} (\overline{E}-\overline{\alpha}_{s'})}{\prod_{s' \ne s} (\overline{\alpha}_s-\overline{\alpha}_{s'})}.
\]
Clearly, $E^i{}_kP[s]^k{}_j = \alpha_s P[s]^i{}_j $ and
$\overline{P}[s]_i{}^k \overline{E}_k{}^j = \overline{\alpha}_s \overline{P}[s]_i{}^j$. If $W=V_0(\lambda)$ with highest weight $\lambda$, by construction \cite{green1971ci} $T[s]$ (or $\overline{T}{[}s{]}$) only has components in the $gl(n)$ module with highest weight $\lambda+\delta_s$, where $\delta_s$ is a weight of $\Lambda_1^*$ (or $\Lambda_1$, respectively). 
An important aspect of the polynomial identities is the following \cite{green1971ci}. Although generically the identity is of degree $n$, in practice an identity of \emph{reduced} degree may hold. If $\delta_s$ is a weight of $V_0(\Lambda_1^*)$, then the factor with root $\alpha_s$ is retained only if $\lambda+\delta_s$ \emph{is} a dominant integral weight. Correspondingly, the projectors $P{[}s{]}$, $\overline{P}{[}s{]}$ for non-dominant weights $\lambda+\delta_s$ vanish identically, and the decomposition of $T_i$ is \emph{only} over the remaining nonzero shifts. 
With these enveloping algebra methods in hand we turn to the analysis of the structure of the Kac modeules for $gl_2(n/1)$.


\subsection{Shift operator analysis and atypicality conditions for $gl_2(n/1)$}
\label{subsec:AtypicalityConditions}
We return to (\ref{eq:ExplicitForm}) following (\ref{eq:GeneralForm}), Theorem \ref{thm:ModifiedKac}. Consider the first level, the states
$Q_i \overline{S} \otimes v$ for $v\in V_0(\Lambda')$ and $i=1,2,\cdots, n$. It is evident from 
(\ref{eq:VectorOpRelations}) above that their span is isomorphic to the $gl(n)$-module 
$V_0(\Lambda_1^*) \!\otimes\! V_0(\Lambda')$, the tensor product of the starting module $V_0(\Lambda')$ with the contragredient vector representation, with highest weight 
$\Lambda_1^* = (0,0,\cdots,0,-1)$. The following analysis relies on the 
resolution of the structure of this tensor product as a direct sum of irreducibles.
Abstractly, this is of course done by inserting the corresponding projection operators; however, it is clear from 
\S \ref{subsec:PolynomialIdentities} above that we can identify the $Q_i$ formally as the components of a vector operator $Q$ which admits a covariant decomposition as a sum of shift components $Q = \sum_r Q{[}r{]}$, each of which is also a vector operator, but which maps to states of only one specific irreducible summand.

\mybenv{Theorem} \textbf{First level atypicality conditions for $gl_2(n/1)$}:\\
\label{thm:FstLevelAtyp}
Let $\delta_s$ be a weight of $V_0(\Lambda_1^*)$. The irreducible Kac module $V(\Lambda)$ contains the $gl(n)$ submodule $V_0(\Lambda+ \delta_s)$ iff $\Lambda + \delta_s$  is a dominant integral weight, and $a_s(\Lambda_1,\Lambda_2,\cdots, \Lambda_n) \ne 0$ for a certain polynomial function in the highest weight labels.\\

\noindent
\textbf{Proof}:\\
We follow the method used in Gould \cite{Gould1989ar} for the Lie superalgebra $gl(n/1)$. Apply the tensor projection method outlined above to establish that the level one subspace $\left< Q{}_i \overline{S}\otimes v \right> $ is the direct sum of projected parts $\left<Q[s]{}_i \overline{S}\otimes v \right>$. Using the definitions and rearrangement identities, and the tensor product $\otimes_{U_-}$, we manipulate the spanning states as follows:
\begin{align}
\label{eq:ArCoeff}
Q[s]{}_i \overline{S}\otimes v = & \, \overline{P}[s]_i{}^k Q_k \overline{S} \otimes v
=  \overline{P}[s]_i{}^k \big(\overline{S}_k A^k{}_i \big) \otimes v \nonumber \\
\equiv  & \, \big(\overline{S}_k A^k{}_\ell \big) P{[}s{]}^\ell{}_i \otimes v 
\equiv \big(\overline{S}_k A^k{}_\ell \big)  \otimes P{[}s{]}^\ell{}_i v
\equiv \overline{S}_k \otimes A^k{}_\ell P{[}s{]}^\ell{}_i v \nonumber \\
\equiv & \, a(\alpha'_s, C'_1, C'_2)\, \overline{S}_k \otimes P{[}s{]}^k{}_i v, 
\end{align}
where in the last step, thanks to the characteristic identity, and the properties of the projectors noted above, $E^i{}_kP[s]^k{}_j = \alpha_s P[s]^i{}_j $ and
$\overline{P}[s]_i{}^k \overline{E}_k{}^j = \overline{\alpha}_s \overline{P}[s]_i{}^j$, the operator  $A^i{}_j$ reduces to the form $a(\alpha'_s, C'_1, C'_2)\delta^i{}_j$ by substituting the root $\alpha'_s \delta^i{}_j$ for 
$E^i{}_j$, with $a(\alpha'_s, C'_1, C'_2)$ the corresponding polynomial in $\alpha'_s=\Lambda_s-1+n-s$, and the Casimir eigenvalues $C'_1$, $C'_2$ in the representation $V_0(\Lambda')$ carried by the states $v$. Denote the corresponding polynomial of the components of $\Lambda = \Lambda' + 2 \rho_1$ by $a_s(\Lambda) \equiv a_s(\Lambda_1,\Lambda_2,\cdots, \Lambda_n)$. \\ \mbox{} \myeenv
For the second level states, the projected parts of the antisymmetric tensor
$Q_iQ_j \overline{S}$ acting on $V_0(\Lambda')$ are needed. The decomposition is known to be \cite{Gould1989ar,GouldBrackenHughes1989br}
$Q[r,s]{}_{ij} = Q[r]_iQ[s]_j + Q[s]_iQ[r]_j$ for each weight $\delta_r+\delta_s$ of $V_0(\Lambda_2^*)$, $1\le r<s\le n$.  Rearrangement of the tensor projections acting on $B^{kl}{}_{ij}$ leads to a formulation equivalent to (\ref{eq:ArCoeff}), giving criteria for the presence of 
the contributing even modules. For the purposes of this paper we give in the 
appendix, \S \ref{subsec:TensorProj}, a summary of the combinatorial manipulations which would be needed explicitly for this (and higher) levels, and provide a conjecture (Theorem \ref{thm:Conjecture} ) on the nature of polynomials 
$a_{rs}(\Lambda_1,\Lambda_2,\cdots, \Lambda_n)$, ($a_{rst}(\Lambda_1,\Lambda_2,\cdots, \Lambda_n)$, $\cdots$ ) governing the structure of atypicality conditions in such cases.

Although we have not given the complete structure of atypical modules for $gl_2(n/1)$ in closed form, the general approach is clear. The general level one result, Theorem \ref{thm:FstLevelAtyp}, demonstrates the combinatorial complication of the quadratic case. A unique feature which does not exist in the same way for Lie superalgebras is the emergence of classes of \emph{truncated} atypical irreps wherein \emph{all} potential even irreducible modules at a given level are annihilated as a result of polynomial identities satisfied by the even generators. For $gl_2(n/1)$, nontrivial cases associated with quadratic identities can be sought at level one and two, denoted zero- and one- step atypicals respectively (Theorem \ref{thm:ZeroOneStep}). To finish our examination of $gl_2(n/1)$ we turn to a more systematic examination of these cases. In the same vein we give a complete formulation of atypicality criteria at level 1 (Corollary \ref{thm:FirstLevelAtyp}) for a class of Kac modules $V(\Lambda)$ for which the even generators satisfy a quadratic characteristic identity on the top even module $V_0(\Lambda)$.




\subsection{Classes of atypical modules for $gl_2(n/1)$}
\label{subsec:ExamplesForGL21}

The above results are complete but rather implicit. We have not given $a_s$ (or equivalents such as $a_{rs}$ at higher levels, \S \ref{subsec:TensorProj}) explicitly as polynomials in the components of the highest weight vector $\Lambda$, as the expressions are tedious and non-transparent in this form. We finish this section with some examples of atypical modules, for the family of Kac modules with $gl(n)$ highest weight $\Lambda = (\mu,\mu,\cdots,\mu;\nu,\nu,\cdots,\nu)$ $\equiv (\mu^r, \nu^{n\!-\!r})$ in Cartesian components, where $\mu-\nu \in {\mathbb Z}^+$ and $r=1,2,\cdots, n-1$. An alternative parametrisation is $\mu = w + k$, $\nu = w$ where $k \in {\mathbb Z}^+$, which makes it clear that the $gl(n)$ modules $V_0(\Lambda)$ that we consider are rational (tensor density) representations, equivalent to the direct product of a finite-dimensional tensor representation with highest weight $(k,k,\cdots k, 0,\cdots,0) \equiv (k^r,0^{n-r})$ (represented by a rectangular $r\! \times\!k$ Young diagram), with the one dimensional character $det^w$. Define 
\[
\overline{\mu}:= \mu+n-r, \quad \overline{\nu} := \nu, \quad \mbox{for} \quad r=1,2,\cdots,n-1. 
\]
The characteristic identity in this case is quadratic, so that there are generically only two projections at each level, only two roots to deal with, and simple expressions can be given for the $a_s$, and related polynomials. The case $r=n$ (or equivalently $r=0$) requires a separate discussion as the characteristic identity becomes linear (with only one root). Firstly we collect some relevant data about the characteristic identity and
invariants. 
\mybenv{Lemma} \textbf{Characteristic identity and Casimir eigenvalues for $\Lambda = (\mu^r, \nu^{n\!-\!r})$}:\\
Consider the class of Kac modules of $gl_2(n/1)$ with $\Lambda$ as above. In terms of the quantities 
$\overline{\mu}$, $\overline{\nu}$ defined above, we
have the following data for operators acting on states 
$v \in V_0(\Lambda')$:
\begin{description}
\item[(i)] The Gel'fand array satisfies the quadratic matrix identity
\begin{align}
\label{eq:QuadId}
E^2 - (\overline{\mu} + \overline{\nu}-2) E + (\overline{\mu}-1) (\overline{\nu}-1) = & \, 0.
\end{align}
\item[(ii)] The linear and quadratic Casimir invariants are
\begin{align}
\label{eq:Casimirs}
C_1' = & \, r \overline{\mu} + (n-r)\overline{\nu} -r(n-r)-n, \qquad 
C_2' = (\overline{\mu}+\overline{\nu}-2)C_1' -n(\overline{\mu}-1)(\overline{\nu}-1).
\end{align}
\item[(iii)] The adjoint operator $A$ takes the form
\begin{align}
\label{eq:Aform}
\! \! \! \! \! \!  A^i{}_j =  -\big[(r-1) \overline{\mu} + (n-r-1)\overline{\nu} -r(n-r)\big] E^i{}_j + & \nonumber \\
  + \big[ \texttt{c} \!-\!n\!+\!1 \!+\! (\textstyle{\frac 12}n\!-\!1)(\overline{\mu}\!-\!1)(\overline{\nu}\!-\!1)
\!+\! \textstyle{\frac 12}\big(r\overline{\mu} \!+\! (n\!-\!r)(\overline{\nu} \!-\!r)\!-\!n\big)\big((r\!-\!1) & \overline{\mu} \!+\! (n\!-\!r\!-\!	1)\overline{\nu} \!-\!r(n\!-\!r)\!-\!1\big) \big] \delta^i{}_j . \qquad \qquad \qquad \qquad \mbox{}
\nonumber \\
&
\end{align}
\item[(iv)] The operator $B$ takes the form
\begin{align}
\label{eq:Bform}
B^{k \ell}{}_{ij} = & \,\big[(r\!-\!1) \overline{\mu} + (n\!-\!r\!-\!1)\overline{\nu} \!-\!r(n\!-\!r)\!-\!1\big](E\delta)^{k \ell}{}_{ij} - 
 \big[r\overline{\mu} \!+\! (n\!-\!r)\overline{\nu} \!-\!r(n\!-\!r)\!-\!3\big](\delta E)^{k \ell}{}_{ij} + 
\qquad \qquad \mbox{} \nonumber \\
\!+\!& \, \big[ \texttt{c} \!-\!n\!+\!1 \!+\! \textstyle{\frac 12}(n\!-\!1)(\overline{\mu}\!-\!1)(\overline{\nu}\!-\!1)
+   \nonumber \\
& \, \qquad \quad +  \textstyle{\frac 12}\big(r\overline{\mu} \!+\! (n\!-\!r)\overline{\nu} \!-\!r(n\!-\!r)\!-\!n\big)\big((r\!-\!1) \overline{\mu} \!+\! (n\!-\!r\!-\!1)\overline{\nu} \!-\!r(n\!-\!r)\!-\!1\big) \big](\delta \delta)^{k \ell}{}_{ij}, 
\qquad \mbox{} \nonumber \\
& 
\end{align}
where we define the antisymmetric tensor combination $(XY)^{k \ell}{}_{ij} :=
\big( X^k{}_i Y^\ell{}_j - X^\ell{}_i Y^k{}_j \big)$ for adjoint operators $X^k{}_i$, $Y^\ell{}_j$ (including $\delta^k{}_i$).
\end{description}
\noindent
\textbf{Proof}:
\begin{description}
\item[(i),(ii)]:\\
 $\overline{\mu}$, $\overline{\nu}$  are by definition the roots of the quadratic identity satisfied by $E$. $C_1'$ is simply the sum $\sum_{r=1}^n \Lambda'_r$, while 
$C_2'$ follows by taking the matrix trace of the characteristic identity (or directly from the standard expression for the Casimir invariant in terms of the highest weight labels).
\item[(iii),(iv)]:\\
Directly from (\ref{eq:AsAndBs}), substituting (\ref{eq:QuadId}) and (\ref{eq:Casimirs}). Again $\langle E \rangle = C_1'$,
$\langle E^2 \rangle = C_2'$ appropriate to $V_0(\Lambda')$. Schematically let us express the quantities $A$, $B$ from (\ref{eq:AsAndBs}) as
\begin{align}
A^i{}_j = & \, (E^2)^i{}_j - a_1 E^i{}_j + a_0, \nonumber \\
B^{ij}{}_{k\ell} = & \, (E^2 \delta)^{ij}{}_{k\ell} -b_1(E \delta)^{ij}{}_{k\ell} -\overline{b}_1(\delta E)^{ij}{}_{k\ell}
                                            +b_0(\delta \delta)^{ij}{}_{k\ell}\,,
\label{eq:Schematics}
\end{align}
where 
\begin{align}
\label{eq:Coeffs}
a_1 = & \, C_1'+n-2 \qquad a_0 = \texttt{c}-(n\!-\!1)-\textstyle{\frac 12}(C_2'-(C_1')^2-(n\!-\!3)C_1'), \nonumber \\
b_1 = & \, a_1-1, \qquad \overline{b}_1=-1, \qquad b_0=\texttt{c}-(n\!-\!2)-\textstyle{\frac 12}(C_2'-(C_1')^2-(n\!-\!5)C_1').
\end{align}
Finally substituting
\begin{align}
(E^2)^i{}_j  = & \, s' E^i{}_j  - p' \delta^i{}_j
\label{eq:QuadraticId}
\end{align}
where $s' := \overline{\mu}+\overline{\nu}-2$, $p'= (\overline{\mu}-1)(\overline{\nu}-1)$ as above leads to 
\begin{align}
\label{eq:AformBform}
A^i{}_j = & \, (s'- a_1) E^i{}_j + (a_0-p'), \nonumber \\   
B^{k\ell}{}_{ij}  = & \, (s'-b_1)(E \delta)^{ij}{}_{k\ell} -\overline{b}_1(\delta E)^{ij}{}_{k\ell} +(b_0-p')(\delta \delta)^{ij}{}_{k\ell}
\end{align}
whose coefficients are expanded in full in (iii), (iv).
\end{description}
\mbox{} \myeenv
\mybenv{Corollary} \label{thm:FirstLevelAtyp} \textbf{First level atypicality conditions for $gl_2(n/1)$ Kac modules $V(\mu^r,\nu^{n-r})$}:\\
Let $a_r(\overline{\mu}, \overline{\nu}){\delta^i}_j$, $a_n(\overline{\mu}, \overline{\nu}){\delta^i}_j$ be the polynomials arising by replacing ${E^i}_j$ by $(\overline{\mu}\!-\!1){\delta^i}_j$, $(\overline{\nu}\!-\!1){\delta^i}_j$, respectively, in the expression for the effective form of ${A^i}_j$ above. Then the atypicality conditions for the $gl_2(n/1)$-module $V(\mu^r,\nu^{n\!-\!r})$ are: 
$a_r(\overline{\mu}, \overline{\nu})=0$ and $a_n(\overline{\mu}, \overline{\nu})=0$.\\

\noindent
\textbf{Proof}:\\
From Theorem \ref{thm:FstLevelAtyp} (see (\ref{eq:ArCoeff})), atypicality is signalled by the vanishing of $P[s]_i{}^k A_k{}^j \equiv a_s P[s]_i{}^j$ for each root $\delta_s$ of $V_0(\Lambda_1^*)$ such that $\Lambda'+\delta_s$ is dominant integral. In this case the only choices are $s=r$, $s=n$, for which $P[r]_i{}^k E_k{}^j= (\overline{\mu}\!-\!1)P[r]_i{}^j$ and  $P[n]_i{}^k E_k{}^j= (\overline{\nu}\!-\!1)P[n]_i{}^j$, respectively.
\mbox{} \myeenv
 
\mybenv{Theorem} \label{thm:ZeroOneStep}\textbf{Zero- and one-step atypical $gl_2(n/1)$ Kac modules $V(\mu^r,\nu^{n-r})$}:
	\begin{description}
	\item[(i)] The Kac module $V(\mu^n)$ is of zero-step type (level 0 even submodule only, $V(\mu^n) \cong V_0(\mu^n)$), if the following condition holds ($n \ge 3)$:
\begin{align}
\label{eq:ZeroStep1D}
\left({\mu} -\frac{\frac 12n}{n\!-\!2}\right)^2 = & \,
	\left(\frac{\frac 12n}{n\!-\!2}\right)^2 - \frac{2\texttt{c}}{(n\!-\!1)(n\!-\!2)}. 
\end{align}
	\item[(ii)]
	The Kac module $V(\mu^r,\nu^{n-r})$, $r=1,2,\cdots,n$ is of zero-step type (level 0 even submodule only, $V(\mu^r,\nu^{n-r})\cong V_0(\mu^r,\nu^{n-r})$), if the following conditions hold ($n \ge 3)$:
	\begin{align}
	\label{eq:ZeroStepExplicit}
	(r-1)\overline{\mu} + (n-r-1)\overline{\nu}  = & \, r(n-r), \nonumber \\
	\mbox{and} \qquad \left(\overline{\mu} - \frac{n\!+\!1}{n}\right)\left(\overline{\nu} - \frac{n\!+\!1}{n}\right) = & \,
	\left(\frac{n\!+\!1}{n}\right)^2 - \frac{2\texttt{c}\!+\!1}{n}. 
	\end{align}
	\item[(iii)]
	For highest weight $\Lambda = (\mu^r,\nu^{n\!-\!r})$ there are no $gl_2(n/1)$ atypical modules of one-step type (level 1 and level 2 even submodules only) for $n \ge 3$.
	\end{description}
\textbf{Proof}:
\begin{description}
\item[(i)] 
This case cannot be obtained from case (ii) as the characteristic identity is linear -- that is, the module $V_0(\Lambda')$ is one dimensional with $E^i{}_j = \mu'\delta^i{}_j$, $\mu'=\mu-1$, and so $(E^2)^i{}_j = \mu'{}^2\delta^i{}_j$, $C_1'=n\mu'$, $C_2'=n\mu'{}^2$. Making these substitutions directly into $A^i{}_j$ given in (\ref{eq:Aform}) leads to (\ref{eq:ZeroStep1D}). 
\item[(ii)]
From Theorem \ref{thm:FstLevelAtyp} and (\ref{eq:ArCoeff}), $V(\mu^r,\nu^{n-r})\cong V_0(\mu^r,\nu^{n-r})$ entails the simultaneous satisfaction of \emph{both} level one atypicality conditions, namely $a_r(\overline{\mu}, \overline{\nu}) = a_s(\overline{\mu}, \overline{\nu})=0$.  If $\overline{\mu}\ne \overline{\nu}$ the situation is equivalent to the vanishing of the adjoint operator $A^i{}_j$ on $V_0(\Lambda')$ in the induced module construction. From (\ref{eq:Aform}), the numerical coefficients of both $E^i{}_j$ and $\delta^i{}_j$ must vanish. The second of (\ref{eq:ZeroStepExplicit}) arises from the vanishing of the $\delta^i{}_j$ coefficient by use of the first condition, and rearranging (assuming $n \ge 3$). 

We give an alternative
demonstration of the existence of zero-step modules in a slightly more
general context. Although derived from the vanishing of
$Q_i\overline{S} \otimes v \equiv \overline{S}_j \otimes A^j{}_iv$, that is, the
vanishing of the operator $A^j{}_i$ acting on the states $v$ belonging
to $V_0(\Lambda')$, the outcome is of course that the atypical Kac module
$V(\Lambda)$ is in fact isomorphic to $V_0(\Lambda)$ itself, with the
odd generators represented as identically zero. This can only be the case
if the anticommutator $\{\overline{Q}^i, Q_j\} = C^i{}_j$ is such that
$C^i{}_j$ vanishes on $V_0(\Lambda)$.

We can confirm this in a general way as follows. Let the quadratic
characteristic identity be $0=Q'{}^i{}_j = (E^2){}^i{}_j - s' E^i{}_j
+p'\delta^i{}_j$ acting on $V_0(\Lambda')$. Here $s'$, $p'$ are the sum and
product of the roots, respectively. Identifying $Q'{}^i{}_j$ and $A^i{}_j$
we have 
\[
s'= C_1' + n-2,
\qquad
p' = \texttt{c}-n+1 -\textstyle{\frac 12}\big( C_2' 
- C_1'{}^2 -(n-3)C_1').
\]
On the other hand, on $V_0(\Lambda)$ itself we have
$C_1 = C_1' +n$,
$C_2 = C_2' +2C_1'+n$, 
the characteristic identity entails the sum and product $s'+2$, $p'+s'+1$,
respectively. With these interrelations it is easily shown that, if the 
identification between $Q'$ and $A$ (acting on $V_0(\Lambda')$) is made as above, then the appropriately adjusted quadratic identity $Q$, and the anticommutator bracket $C$ (as given in (\ref{eq:gln1StructConsts}), are indeed also identical on
$V_0(\Lambda)$. Thus at least for the zero-step atypical case, the constraints are available directly from the structure relations without the need for the induced module construction.
\item[(ii)]
From the discussion following Theorem \ref{thm:FstLevelAtyp}, one-step atypical representations require the vanishing of \emph{all}
tensor projections of the states $Q_iQ_m \overline{S}\otimes v$, that is, we require
\begin{align}
0=& \, Q_iQ_m \overline{S}\otimes v = Q_i \overline{S}_j A^j{}_m\otimes v + Q_i \overline{S} Q_m \otimes v
\nonumber \\
= & \, \big( \overline{S}_{k\ell} B^{k\ell}{}_{ij} + \overline{S}_j Q_i \big)A^j{}_m \otimes v \nonumber \\
\equiv & \, \overline{S}_{k\ell} B^{k\ell}{}_{ij} A^j{}_m \otimes v ,
\end{align}
where in the first line we have used $\overline{S}Q_m\otimes v = \overline{S}\otimes Q_m v \equiv 0$, and in the second line
$\overline{S}_j \otimes Q_i A^j{}_m v = \overline{S}_j \otimes {[}Q_i ,A^j{}_m{]} v\equiv \overline{S}_j \otimes C_{im}{}^{jk}Q_k v =0$ for some even elements $C_{im}{}^{jk}$. Thus we require simply $B^{k\ell}{}_{ij} A^j{}_m  v = 0$.

Using the explicit expressions for the forms of the tensors $B$ and $A$ given earlier, and exploiting the quadratic characteristic identity for $E$, this condition can be reduced to a set of numerical constraints on certain operator coefficients as follows. Recall the schematic notation introduced previously,
\begin{align}
\label{eq:Schematics}
A^i{}_j = & \, (E^2)^i{}_j - a_1 E^i{}_j + a_0, \nonumber \\
B^{ij}{}_{k\ell} = & \, (E^2 \delta)^{ij}{}_{k\ell} -b_1(E \delta)^{ij}{}_{k\ell} -\overline{b}_1(\delta E)^{ij}{}_{k\ell}
                                            +b_0(\delta \delta)^{ij}{}_{k\ell}, \nonumber \\
(E^2)^i{}_j  = & \, s' E^i{}_j  - p' \delta^i{}_j.
\end{align}
Expanding and using the quadratic identity leads to
\begin{align}
A^i{}_j = & \, (s'- a_1) E^i{}_j + (a_0-p'), \nonumber \\                                   
B^{k\ell}{}_{ij} A^j{}_m = & \,
(s'\!-\!a_1)(s'\!-\!b_1)(E E)^{k\ell}{}_{ij} + (a_0\!-\!p')(s'\!-\!b_1) (E \delta )^{k\ell}{}_{ij} + \nonumber \\
 & \, \big((s'\!-\!a_1)(b_0+s'\!-\!p')\!-\!(a_0\!-\!p')\overline{b}_1  \big)(\delta E )^{k\ell}{}_{ij} 
+ \big(  (a_0\!-\!p')(b_0\!-\!p') + p'(s'\!-\!a_1)   \big)(\delta \delta)^{k\ell}{}_{ij}. \nonumber
\end{align}
Taking the vanishing of the numerical coefficients in turn, for the $(EE)$ term we have either $s'=a_1$ or $s'=b_1$. Setting $s'=a_1$ means $b_1=-1$ and so the second term requires $a_0\!-\!p' =0$, and we trivially recover the zero-step case. If $s'=b_1$, the $(\delta E)$ coefficient becomes $a_0-b_0=s'$ or, $C_1'-1 = C_1'+n-3$ which leads to the degenerate Lie superalgebra case $n=2$.  
\end{description}
\mbox{} \myeenv

\begin{table}[tbc]
  \centering 
\begin{tabular}{| l || lllll lllll ll ||}
\hline
$n$ & 3 & 4&  5&    6&     7&  7&  7&  8& 9& 9& 10& 10\\
$r$ &  2  & 2&  2&    2&    2&  3&  4&  2& 2& 5&   2& 4\\
$k$ & 1  & 2&  3&    4&     5&  2& 1&   6& 7& 1&  8& 2 \\
\hline
\end{tabular}
\caption{Zero-step atypicals: $gl_2(n/1)$ Kac modules $V(k^r,0 ^{n-r})$ 
for $n=3,4,\cdots,10$ }
\label{tab:ZeroStepExx}
\end{table}

As a final illustration we give in Table \ref{tab:ZeroStepExx} concrete instances of zero-step atypical modules for various low-dimensional cases. For these purposes assume $w=0$, that is, we take $\mu=k$ and $\nu=0$ for true tensor representations. Further, we can regard the second of (\ref{eq:ZeroStepExplicit}) as the defining condition for the central charge $\texttt{c}$ given the choice of other labels, and so we present solutions to (\ref{eq:ZeroStepExplicit}$a$), namely $(r-1)(k+n-r)=r(n-r)$, $r=2,\cdots, n-1$ (see Table \ref{tab:ZeroStepExx}). More generally it can be seen from both parts of (\ref{eq:ZeroStepExplicit}) that simultaneous solutions may exist for $\mu$ and $\nu$ for arbitrary fixed parameters $n,r,\texttt{c}$, but that the requirement $\mu-\nu \in {\mathbb Z}^+$ is a severe restriction.

%


\section{Discussion and Conclusions}
\label{sec:Examples}
 In this paper we have developed the structure and
representation theory of a distinguished class of ${\mathbb Z}_2$-graded
polynomial algebras of degree 2, which we have termed quadratic
Lie superalgebras. Their defining relations and PBW property have been
formulated and proven along the same lines as for Lie superalgebras (but
using the property of Koszulness in the quadratic case). The `type I'
class includes several examples studied earlier inspired by various
physical contexts. Some general observations on the structure of the
structure constants and the generalized Jacobi identities have been
made, although a classification is not feasible with present methods.
Given the PBW Theorem for the enveloping algebra, the balanced type I
case admits a formulation of induced modules following Kac and
especially the modified construction of Gould. A specific case, the
quadratic modification of the simple Lie superalgebra $sl(n/1)$, denoted
here $gl_2(n/1)$, has been considered in detail, and tensor projection
techniques used to expose the structure of level 1 atypicality
conditions. An outline of the application of these methods has been provided 
for the analysis of induced modules at level $\ge 2$, and a conjecture given for 
atypicality conditions at level 2. 

A new feature not present for the Lie superalgebra case but which
distinguishes the quadratic (and presumably all polynomial cases) is the
existence of truncated atypicals, where all states of the Kac module at
a certain level can vanish due to polynomial identities satisfied by the
even generators at a given level. These have been investigated at level
one and two for a class of Kac modules of $gl_2(n/1)$ where the even
Gel'fand generators satisfy a quadratic polynomial identity; the
corresponding truncated modules are denoted as zero-step and one-step
modules, respectively. It has been found that zero-step modules are a
generic feature of this class, whereas there are in fact no one-step
modules. This is in contrast to $gl(n/1)$. In that case,
appropriate graded tensor products of the defining representation lead
to atypical modules with a simple combinatorial structure, for example
with decomposition into $gl(n)$ of the form $k\Lambda_1 + (k-1)\Lambda_1$ in highest weight  notation (or $\{k\} + \{k-1\}$ in
partition notation).

Let us recall the context of constructions of quadratic superalgebras arising from the study of gauge invariant fields in Hamiltonian lattice QCD \cite{JarvisRudolph2003,JarvisKijowskiRudolph2005} and the relevance of atypicality conditions therein.
Examples of composite operators arising in those studies were the baryonic type 
fields $\psi_{ijk}= \varepsilon^{abc}c_{ai}c_{bj}c_{ck}$ and $\overline{\psi}{}^{ijk}= 
\varepsilon^{abc}c{}^\dagger_{ai}c^\dagger_{bj}c^\dagger_{ck}$, where $i,j,k = 1,2,\cdots, n$ are spatial lattice, spin and flavour labels, and $a,b,c=1,2,3$ are colour labels, for fermionic creation and annihilation operators $c,c^\dagger$. On the other hand, a related construction of gauge invariant fermions $a,a^\dagger$ led to the colourless composites $Q_{ijk}=a_ia_ja_k$,
$\overline{Q}{}^{ijk} = a^\dagger_i a^\dagger_ja^\dagger_k$. Clearly, the objects $\psi_{ijk}$, $\overline{\psi}{}^{ijk}$ and $Q_{ijk}$,
$\overline{Q}{}^{ijk}$ have anticommutator brackets quadratic in the corresponding even $gl(n)$ generators $E^i{}_j = c^\dagger_{ai}c_{aj}$ and $a^\dagger_i a_j$, respectively, yielding instances of the class I type quadratic superalgebras with odd generators in the non-fundamental representations with highest weights $3\Lambda_1$ and $\Lambda_3$, respectively -- that is, the quadratic Lie superagebras 
$gl_2(n/3\Lambda_1\!+\!3\Lambda_1^*)$ and $gl_2(n/\Lambda_3\!+\!\Lambda_3^*)$ 
(see also \S \ref{sec:Atypicals}, introduction). 

The role of atypical and zero-step modules can be appreciated when the action of the $Q$, $\overline{Q}$ generators in the usual fermionic Fock space is considered. Take for instance $n=4$ (the case $gl_2(4/\Lambda_3+\Lambda_3^*)$) and 
consider the 6 occupation number 2 states $a^\dagger_m a^\dagger_n |0\rangle$ (corresponding to the $gl(4)$-module $\Lambda_2$).
Clearly by normal ordering, the ${Q}_{ijk}$ annihilate these states, and equally so do the $\overline{Q}{}^{ijk}$ -- there are no 5-particle states. This situation can be explained abstractly as follows\footnote{
See \cite{JarvisRudolph2003,JarvisKijowskiRudolph2005} for comparable explicit calculations.}.
Firstly, note that the anticommutator bracket $\{Q_{ijk}, \overline{Q}{}^{pqr}\}$ can be displayed in the form of a polynomial in the standard generators ${\mathcal E}^{ijk}{}_{pqr}$ which are the matrix elements of the object (\ref{eq:CutCoproduct}) in $End(V_0(\Lambda_3^*)\otimes End(V_0(\Lambda))$, namely
$\{Q, \overline{Q}{} \} =  -\textstyle{\frac 32} \big({\mathcal E}^2 - (n+3-N) {\mathcal E} + 4)$,
where ${\mathcal E}^{ijk}{}_{pqr} := \frac 16 (E^i{}_p \delta^j{}_q \delta^k{}_r \pm \cdots )$ ($3^2\times 2=18$ terms),
the unit is the corresponding antisymmetric tensor ${\delta}^{k\ell m}{}_{pqr} := \frac 16 (\delta^k{}_p \delta^\ell{}_q \delta^m{}_r \pm \cdots )$ ($3!=6$ terms), and $N$ is the number operator $\sum a^\dagger_i a_i$. The roots of the characteristic identity for ${\mathcal E}$ correspond to weights $\delta$ in $V_0(\Lambda^*_3)$ such that $\Lambda +\delta$ is dominant integral. In this case for $n=4$, $N=2$, $\Lambda=\Lambda_2$ and in Cartesian components $\Lambda_2 = (1,1,0,0)$, $\Lambda^*_3 = (0,-1,-1,-1)$ and the appropriate weights $\delta$ are $(0,-1,-1,-1)$ and $(-1,0,-1,-1)$. From (\ref{eq:CutCoproduct}), the corresponding roots are 1 and 4 respectively (see also \cite{green1971ci}), and we recognise that for this case (as $n\!+\!3\!-\!N=5$), the occupation number 2 states indeed form a zero-step atypical representation of the associated quadratic Lie superalgebra, $gl_2(4/\Lambda_3+\Lambda_3^*)$. Similar considerations apply for other low dimensional cases of Fock space constructions of such superalgebras.

In conclusion we note that a considerable literature exists on the subject of (homogeneous) algebras and superalgebras which could be drawn upon to elucidate some of the representation-theoretical questions encountered in this study.
We cite in particular \cite{HaiKriegkLorenz2008,HaiLorenz2007} for homological aspects, and \cite{DuboisVioletteLandi2010} for constructions related to free differential algebras (see also \cite{Priddy1970kr}).

\subsection*{Acknowledgements}
PDJ and GR thank the Department of Theoretical Physics, University of Leipzig, and the School of Mathematics and Physics, University of Tasmania for hospitality and support during collaborative visits. We wish to express our sincere appreciation for  grants from the Alexander von Humboldt Foundation which made these visits possible. LAY acknowledges the support of a Commonwealth postgraduate award. Part of this work was done when PDJ was visiting the Department of Sciences, Division of Mathematics, Technical University of Crete, and the hospitality of this institute and colleagues there is gratefully acknowledged. Similar appreciation is expressed to the Fulbright Foundation, and the Department of Statistics, University of California Berkeley and colleagues, for a visit as an Australian senior Fulbright scholar. 
PDJ also acknowledges an earlier Erskine visiting Fellowship at the Department of Physics, University of Canterbury, where part of this work was initiated, and thanks William Joyce for discussions. The authors thank the anonymous referees for detailed feedback which has significantly improved the paper, and for drawing our attention to \cite{HaiKriegkLorenz2008,HaiLorenz2007}. We thank the authors of \cite{DuboisVioletteLandi2010} for providing us with a copy of their paper.







%



\hfill
\pagebreak
\begin{appendix}
\setcounter{equation}{0}
\renewcommand{\theequation}{{A}-\arabic{equation}}
\section{Appendix}
\label{sec:appendix}
\subsection{Algebras of PBW type}
\label{subsec:PBWproof}
We present here the required proof of the PBW property of the homogeneous graded quadratic algebra $U(I_2)$ (Lemma \ref {lem:PBWbasis} in the main text) in a slightly more general context than is required for quadratic Lie superalgebras. For the time being we work in any field ${\mathbb K}$ of characteristic zero\footnote{
In their study of nonlinear conformal superalgebras, De Sole and Kac (see for example \cite{DeSoleKac2006}) have noted that the PBW property holds on very general grounds. However, for completeness, and because the method can be adapted to more general cases, we provide here an explicit, constructive proof.}. 
\mybenv{Theorem} \textbf{PBW algebra}:\\
\label{pbw-algebra_thm}
As before let $\{w_a, a=1,\cdots, m+n \}$ be a fixed homogeneous basis for $L$, and consider the defining relations
\[
w_a w_b = (-1)^{|a||b|} w_b w_a +  d_{ab}{}^{cd}w_c w_d
\] 
which generate the ideal
$I_2$. Then the homogeneous quadratic algebra $U(I_2)=T(L)/J(I_2)$ is a PBW-algebra iff an index ordering exists such that only those $d_{ab}{}^{cd}$ are nonvanishing for which both $c$ and $d$ precede $a$ and $b$.

\noindent
\textbf{Proof}:\\
\emph{if}:\\
Fix a total ordering on the index set $\{1,2,\cdots,m+n\}$. Following Priddy \cite{Priddy1970kr} (see also Kr\"{a}hmer \cite{kraehmer2005}), we begin with the distinguished set ${\mathcal S}$ of label pairs, which will control the choice of objects in the enveloping algebra at degree 2 and higher. Generically, for ${\mathcal S}$ are selected all pairs $(a,b)$ for which $w_a w_b + I_2 \notin \big\langle w_c w_d + I_2, (c,d)<(a,b) \big\rangle$, where pairs are compared lexicographically. Then
${\mathcal S} = {\mathcal S}_1$ generates in turn the set ${\mathcal S}_{m\!-\!1} =
\{(a_1,a_2,\cdots,a_m) \mid (a_i,a_{i\!+\!1})\in {\mathcal S}, i=1,\cdots,m\!-\!1  \}$ of chains of $m\!-\!1$ successive admissible pairs, $m\ge 2$. To each $M\in {\mathcal S}_{m\!-\!1}$, is associated the corresponding word $w_M = w_{a_1}w_{a_2}\cdots w_{a_m}$ of length $\ell(M) = m$ in $U(I_2)$.

Now observe from above that in the present case, the defining quadratic relations of $I_2$ may always be written such that the product $w_a w_b$ on the left hand side has $a\geq b$. For the fixed total ordering of $\{1,2,\cdots,m+n\}$, we now make the assumption that both $a$ and $b$ are greater than $c$ and $d$ whenever $d_{ab}{}^{cd}$ is nonzero -- every product of the form $w_aw_b$, $a \ge b$, may be written as a linear sum of products of the form $w_c w_d$, $c\leq d$. Thus these ordered pairs all belong to ${\mathcal S}$. Beginning with the pair $(1,1)$, one may successively add all pairs of the form $(a,b)$ where $a<b$ if $|a|=|b|=1$ and $a\leq b$ otherwise. Hence the set $S_{m\!-\!1}$, $m\geq 2$, simply consists of all ordered $m$-tuples, where here and for the sequel, \emph{ordered} now means that the even (odd) indices are weakly (strictly) increasing. To each such $m$-tuple (word) $M= (a_1,a_2,\cdots,a_m)$ we associate the \emph{ordered monomial} $w_M:=w_{a_1}w_{a_2}\cdots w_{a_m}$.

We adapt the method of Serre \cite{Serre1992lie} to show that these ordered monomials are a linearly independent basis for $U$. Let $V$ be the ${\mathbb K}$-vector space with basis $\{z_M\}_M$ indexed by all admissible ordered words (including $z_\emptyset$ for the empty word), that is, all $m$-chains , $m=0,1,2,\cdots$. For $M=(a_1,a_2,\cdots,a_m)$ and $a<a_1$ (or $a=a_1$ and $w_{a}\in L_0$) we shall say that $a<M$ and let $aM=(a,a_1,a_2,\cdots,a_m)$. Otherwise we write  $a\geq M$. We show that $V$ can be made into a $U(I_2)$-module in such a way that $w_az_M=z_{aM}$ whenever $a\leq M$. Given this, the linear independence of the $w_M$ is easily shown. For it is clear by induction that $w_Mz_{\emptyset}=z_M$, so if $0=\sum c_M w_M$ where $c_M\in {\mathbb K}$, then $0=\sum c_Mw_Mz_{\emptyset}=\sum c_Mz_M$. However, this immediately implies $c_M=0$, because the $z_M$ are linearly independent by assumption.

The remainder of the proof consists in establishing the correct module action on $V$, which we do in the selected basis $\{ w_a\}$. We need to define $w_az_M$ for all $a$ and $M$. We may assume by induction that $w_b z_N$ is defined for all $b$ when $\ell(N)<\ell(M)$, and for $b<a$ when $\ell(N)=\ell(M)$. We assume this has been done in such a way that $w_bz_N$ is a ${\mathbb K}$-linear combination of $z_L$'s with $\ell(L)\leq \ell(N)+1$. We set
\begin{eqnarray}
\label{module_action}
	w_a z_M=
\left\{\begin{array}{lll}%
&z_{aM}, &\mbox{if $a<M$}; \\ 
&(-1)^{|a||b|}w_bw_az_N+d_{ab}{}^{cd}w_cw_dz_N,\quad&  \mbox{if $M=bN$ with $a>b$};\\
&\frac12d_{ab}{}^{cd}w_cw_dz_N, &\mbox{if $M=aN$ and $w_a\in L_{\bar{1}}$}.
\end{array}\right.
\end{eqnarray} 
If either $|a|=0$ or $|b|=0$ then it is clear from the inductive assumptions that the above are well defined since $d_{ab}{}^{cd}=0$. For the case $|a|=|b|=1$ then $(c,d)<(a,b)$ implies that the quadratic term is well defined by inductive assumption also. 
We need only show that 
\begin{eqnarray}
\label{module_condition}
w_aw_bz_N = (-1)^{|a||b|}w_bw_az_N+d_{ab}{}^{cd}w_cw_dz_N
\end{eqnarray}
for all $a,b$ and $N$. Due to the (graded) antisymmetry of (\ref{module_condition}) we may assume $a>b$ and $a\geq b$ for the case $|a|=|b|=1$. By induction on $\ell(N)$ we may assume that 
\begin{eqnarray}
\label{ind_assumption}
w_aw_bz_L=(-1)^{|a||b|}w_bw_az_L+d_{ab}{}^{cd}w_cw_dz_L
\end{eqnarray}	
holds for all $w_a,w_b$ for $\ell(L)<\ell(N)$ (since (\ref{module_condition}) is true for $N=\emptyset$). If $|a|=0$ or $|b|=0$ then 
$d_{ab}^{cd}w_bw_a=0$ and (\ref{module_condition}) is true. If $|a|=|b|=1$ then (\ref{module_condition}) is true for $b<N$, this follows from the second case of the inductive definition (\ref{module_action}), and we need only consider the case $b\geq N$. Let $N=cL$, thus we have $a\geq b\geq c$ and (\ref{module_condition}) becomes
\[
(a,b,c)\quad\quad y_ay_by_cz_L+y_by_ay_cz_L=d_{ab}^{ef}x_ex_fy_cz_L.
\]
We permute the $a,b,c$ cyclically to obtain the equations
\begin{eqnarray*}
&(b,c,a)&\quad\quad y_by_cy_az_L+y_cy_by_az_L=d_{bc}^{ef}x_ex_fy_az_L, \\ 
&(c,a,b)&\quad\quad y_cy_ay_bz_L+y_ay_cy_bz_L=d_{ca}^{ef}x_e x_f y_bz_L.
\end{eqnarray*}
By inspection of the ordering of cases in (\ref{module_action}) we see that equations $(b,c,a)$ and $(c,a,b)$ are true under the inductive assumption; and similarly, by (\ref{ind_assumption}), the right hand side of $(a,b,c)$ becomes
\[
d_{ab}^{ef} x_e x_f y_c z_L=
y_c d_{ab}^{ef}x_e x_f z_L=
y_c(y_ay_b+y_by_a)z_L=(y_cy_ay_b+ y_cy_by_a)z_L.
\]
We apply this to the right hand sides of equations $(b,c,a)$ and $(c,a,b)$, and it is easily checked that $(a,b,c)+(b,c,a)+(c,a,b)$ becomes a trivial identity; hence $(a,b,c)$ is true.\\

\noindent
\emph{only if}: \\
 Let us assume that $d_{ab}^{\hat{c}\hat{d}}$ is non zero for some $|a|=|b|=1$, $|\hat{c}|=|\hat{d}|=0$ where at least one of $\hat{c}$ or $\hat{d}$ is greater than $a$ or $b$. It follows from
\[
y_ay_b= - y_b y_a+ d_{ab}^{cd}x_cx_d
\]
that both $(a,b)$ and $(b,a)$ belong to ${\mathcal S}$, and we shall show that this leads to a linear dependence between the PBW monomials generated by ${\mathcal S}$. We have 
\begin{eqnarray} 
\label{lin_dep}
0&=&(y_ay_a)y_b-y_a(y_ay_b)\nonumber\\
&=&\textstyle{\frac12}d_{aa}^{cd}x_cx_dy_b-y_a(-y_by_a+d_{ab}^{cd}x_cx_d)\nonumber\\
&=&\textstyle{\frac12}d_{aa}^{cd}x_cx_dy_b+y_ay_by_a-d_{ab}^{cd}y_ax_cx_d
\end{eqnarray} 
and it remains to prove that right hand side is a non trivial sum of PBW monomials. It is clear that $(a,b,a)\in {\mathcal S}_3$. The terms of the form $y_ax_cx_d$ and $x_cx_dy_b$ may be reordered in any way due to the commutativity of any even-even or odd-even pair, so one needs only to show that some permutation of $y_ax_cx_d$ is a PBW monomial for all $|a|=1$,  $|c|=|d|=0$. Since $x_cx_d = x_dx_c$ and $d_{ab}^{cd}$ is $c$-$d$-symmetric for all $a,b$ it follows that $(i,j)\in {\mathcal S}$ whenever $i\leq j$ and $|i|=|j|=0$. Similarly, the only relations involving pairs of odd and even indices are $y_ax_c=x_cy_a$, $\forall a,c$, hence $a<c\Rightarrow(a,c)\in S$ and $c<a\Rightarrow(c,a)\in {\mathcal S}$. Therefore at least one permutation of $(a,c,d)$ and $(c,d,b)$ will always belong to ${\mathcal S}_3$. Since none of these are equal to $(a,b,a)$ the right hand side of (\ref{lin_dep}) is a non trivial sum of PBW monomials.

\mbox{} \myeenv

%
%
\mybenv{Lemma}:\\
\label{thm_factor}
Let $A$ be a PBW-algebra. The ordered monomials generated by any total ordering of the basis elements of $X$ are a basis for $A$. \\
%
\noindent
\textbf{Proof}:\\
Let $I$ be an arbitrary ordering of the basis elements in $X$. We may write for example $x_{i_1}<y_{i_2}<x_{i_3}<x_{i_4}<y_{i_5}<...$, where $<$ means $\leq$ when relating two even elements. $I$ induces an ordering $I_0$ on the even subspace, $x_{i_1}<x_{i_3}<x_{i_4}<...$, and similarly $I_1$ on odd subspace, $y_{i_2}<y_{i_5}<...$. 
Let $\hat{I}$ be the total order on $X$ that respects both $I_0$ and $I_1$ and orders all of the even elements before the odd, i.e. $x_{i_1}<x_{i_3}<x_{i_4}<...y_{i_2}<y_{i_5}<...$. By (\ref{pbw-algebra_thm}) the ordered monomials generated by $\hat{I}$ are a PBW-basis for $A$ and each one of these monomials is equivalent, modulo $J(R)$, to a unique monomial ordered by the abitrarily chosen ordering $I$ (pairs of odd elements do not need to exchange places).

\mbox{} \myeenv

(Note in particular that any ordering such that the even elements precede the odd is a satisfactory condition for $A$ to be a PBW-algebra)

\subsection{Generalised Jacobi identities in the quadratic ${\mathbb Z}_2$ graded case}
\label{subsec:GeneralisedJacobi}
It is possible to approach (J1--J3) in component form for a \emph{generic} ${\mathbb Z}_2$-graded quadratic algebra, and then specialise to the quadratic Lie superalgebra case. Thus we take the generators of $I$ to be (dropping the tensor product symbol $\otimes$ for ease of writing) 
\[
w_a w_b - (-1)^{|ab|}w_b w_a - {\texttt d}_{ab}{}^{cd} w_c w_d - {\texttt c}_{ab}^e w_e -
 {\texttt b}_{ab},
\]
with general structure constants ${\texttt b}$, ${\texttt c}$, ${\texttt b}$ and $|ab|=|cd|=|e|$ in the above summations over repeated indices. Thus the generic element of $I_2$ is
\[
w_{ab} = w_a w_b - (-1)^{|ab|}w_b w_a - {\texttt d}_{ab}{}^{cd} w_c w_d
\]
and the projection maps are $\alpha(w_{ab}) = {\texttt c}_{ab}^e w_e$, $\beta(w_{ab})= 
{\texttt b}_{ab}$. 

We identify generic elements of $L\otimes I_2 \cap I_2\otimes L $, by identifying linear combinations of elements of the form $w_a w_{bc}$ with those of the form $w_{ab}w_c$. In the standard Lie superalgebra case, the required basis elements are 
\begin{align}
& \, w_a(w_b w_c \!-\! (\!-\!1)^{|bc|} w_cw_b)+ (\!-\!1)^{|ab|\!+\!|ac|}w_b(w_c w_a \!-\! (\!-\!1)^{|ca|} w_aw_c)+ (\!-\!1)^{|ca|\!+\!|cb|}w_c(w_a w_b \!-\! (\!-\!1)^{|ab|} w_bw_a) \nonumber \\
\equiv & \,  (w_a w_b \!-\! (\!-\!1)^{|ab|} w_bw_a)w_c  + (\!-\!1)^{|ab|\!+\!|ac|}(w_b w_c \!-\! (\!-\!1)^{|bc|} w_cw_b)w_a + (\!-\!1)^{|ca|\!+\!|cb|}(w_c w_a \!-\! (\!-\!1)^{|ca|} w_aw_c)w_b; \nonumber
\end{align}
in the present quadratic case however the $I_2$ basis elements involve the additional terms
${\texttt d}_{ab}{}^{cd} w_c w_d$, and the corresponding cyclic combinations
$w_a(w_b w_c - (-1)^{|bc|} w_cw_b-{\texttt d}_{ab}{}^{cd} w_c w_d) + \cdots$
do not work. Let us temporarily adopt the simplified notation $a\equiv w_a$, $(ab)\equiv 
{\texttt d}_{ab}{}^{cd} w_c w_d$ and 
$(-1)^{|ab|} = {[}{\scriptstyle{ab}}{]}$, $(-1)^{|ab|+|ac|} = \abc$, 
$(-1)^{|ac|+|ad|+ |bc| + |bd|} = \abcd$, and also the abbreviated notation $(ab) \equiv \sum (ab)^{\scriptscriptstyle{1}} (ab)^{\scriptscriptstyle{2}}$ to indicate sums uniquely determined by the index choice ${}_{ab}$. It is possible to add and subtract terms to form the modified equality:
\begin{align}
& \, a(bc - {[}{\scriptstyle{bc}}{]}cb -(bc)) + 
\abc \,b(ca - {[}{\scriptstyle{ca}}{]}ac -(ca)) + 
\cab \,c(ab - {[}{\scriptstyle{ab}}{]}ba -(ab)) \nonumber \\
& \, -\sum \abc (bc)^{\scriptscriptstyle{1}}( a (bc)^{\scriptscriptstyle{2}} -   \abc^{\scriptscriptstyle{2}}(bc)^{\scriptscriptstyle{2}} a)
-\cab (ca)^{\scriptscriptstyle{1}}(  (ca)^{\scriptscriptstyle{2}} b -   \bac^{\scriptscriptstyle{2}} b(ca)^{\scriptscriptstyle{2}})- (ab)^{\scriptscriptstyle{1}}( (ab)^{\scriptscriptstyle{2}} c -  \cab^{\scriptscriptstyle{2}} c (ab)^{\scriptscriptstyle{2}})
\nonumber \\
\equiv & \, (ab - {[}{\scriptstyle{ab}}{]}ba -(ab)) \,c + \abc (bc - {[}{\scriptstyle{bc}}{]}cb -(bc))\,a + 
\cab (ca - {[}{\scriptstyle{ca}}{]}ac -(ca)) \,b
  \nonumber \\
& \, -\sum  (a (bc)^1-\abc^{\scriptscriptstyle{1}}(bc)^1a)(bc)^2 -  
\abc (b (ca)^{\scriptscriptstyle{1}} +\bac^{\scriptscriptstyle{1}}(ca)^{\scriptscriptstyle{1}}b)(ca)^{\scriptscriptstyle{2}} + 
\cab (c (ab)^{\scriptscriptstyle{1}}-\cab^{\scriptscriptstyle{1}}(ab)^{\scriptscriptstyle{1}}c)(ab)^{\scriptscriptstyle{2}}
\label{eq:GenericIntersectI2L}
\end{align}
where the split summands $(\cdot)^{\scriptscriptstyle{1}} \cdot (\cdot)^{\scriptscriptstyle{2}}$ are introduced on each side in anticipation of the type of bracketing required for entries in $L\otimes I_2$ and $I_2 \otimes L$, respectively. Here the additional signs $\bac^{\scriptscriptstyle{1}}$, $\bac^{\scriptscriptstyle{2}}$ are introduced to indicate the switch of $b$ with the first and second factors of $(ca)$, respectively. Sign factors on these summands are reconciled when it is noted that  
$\bac^{\scriptscriptstyle{1}}\cdot \bac^{\scriptscriptstyle{2}}=
\bac$, and by the same token,
$\bac^{\scriptscriptstyle{1}} {[}{\scriptstyle{ab}}{]} = \bac^{\scriptscriptstyle{2}}{[}{\scriptstyle{cb}}{]}$.
If the additional terms still required ${\texttt d}$-type components to qualify as elements of $I_2$, this process would of course continue; but for the quadratic Lie superalgebra case it will turn out that no further terms are needed, and the elements expressed by either side of (\ref{eq:GenericIntersectI2L}) are the spanning set of the intersection
$L\otimes I_2 \cap I_2 \otimes L$. To go further (J1--J3) should be examined for each choice of grading of the triple $abc$ amongst the sectors $\bar{1}\bar{1}\bar{1}$, $\bar{1}\bar{1}\bar{0}$, $\bar{1}\bar{0}\bar{0}$ and $\bar{0}\bar{0}\bar{0}$, and the problem is as formulated in the text.
\subsection{Tensor projections for $V(\Lambda)$ at level $>1$}
\label{subsec:TensorProj}
For the projection
$Q[r,s]{}_{ij} = Q[r]_iQ[s]_j + Q[s]_iQ[r]_j$ we have (compare (\ref{eq:ArCoeff}))
\begin{align}
 \overline{P}{[}r{]}_i{}^k Q_k \overline{P}{[}s{]}_j{}^\ell Q_\ell \overline{S} 
+\overline{P}{[}s{]}_i{}^k Q_k \overline{P}{[}r{]}_j{}^\ell Q_\ell \overline{S}
\equiv & \,
a(\alpha'_s, C') \overline{P}{[}r{]}_i{}^k Q_k \overline{S}_\ell {P}{[}s{]}^\ell{}_j  
+a(\alpha'_r, C')\overline{P}{[}s{]}_i{}^k Q_k \overline{S}_\ell {P}{[}r{]}^\ell{}_j . \nonumber
\end{align}
This expression can be reduced if (\ref{eq:AsAndBs}) is written as 
\begin{align}
\label{eq:Bdecomp}
{[}Q_k, \overline{S}_\ell{]} = & \, \left.\sum\right._B \overline{B}^{(1)}{}_k{}^p \, \overline{S}_{pq} \, {B}^{(2)}{}^{q}{}_\ell \, f_B
\end{align}
where each $\overline{B}{}^{(1)}{}_i{}^k$, ${B}^{(2)}{}^{\ell}{}_j$ is an adjoint operator in the enveloping algebra of $gl(n)$ (a polynomial in $E$ or $\overline{E}$), and the $f_B$ are central elements. Thus
\begin{align}
\big(Q{[}r{]}_i Q{[}s{]}_j + Q{[}s{]}_iQ{[}r{]}_j\big) \overline{S} \otimes v= & \,
\sum\big( \overline{P}{[}r{]}_i{}^k  a(\alpha'_s, C')\overline{B}{}^{(1)}{}_k{}^p \, \overline{S}_{pq} \, {B}^{(2)}{}^{q}{}_\ell \, f_B{P}{[}s{]}^\ell{}_j  + (r\leftrightarrow s)\big) \otimes v\nonumber \\
= 
\left( \sum \big( a(\alpha'_s, C')\right.& \left.\overline{b}{}^{(1)}(\alpha_r, C'')  {b}^{(2)}(\alpha_s,C') f_B(C')\big) \overline{P}{[}r{]}_i{}^k  \, \overline{S}_{k \ell} \,{P}{[}s{]}^\ell{}_j  + (r\leftrightarrow s)
\right) \otimes v \nonumber
\end{align}
where $\overline{b}{}^{(1)}(\alpha_r, C'') \delta_k{}^p$, ${b}^{(2)}(\alpha_s,C')\delta^q{}_\ell$ arise from 
$\overline{B}{}^{(1)}{}_k{}^p$, ${B}^{(2)}{}^{q}{}_\ell$ when contracted with the projection operators 
$\overline{P}{[}r{]}_i{}^k$, ${P}{[}s{]}^\ell{}_j $. Calling the summations $a_{rs}$, $a_{sr}$ respectively, we have
\[
\big(Q{[}r{]}_i Q{[}s{]}_j + Q{[}s{]}_iQ{[}r{]}_j\big) \overline{S} \otimes v = 
a_{rs}(\Lambda) \overline{P}{[}r{]}_i{}^k  \, \overline{S}_{k \ell} \,{P}{[}s{]}^\ell{}_j  + a_{sr}(\Lambda)
\overline{P}{[}s{]}_i{}^k  \, \overline{S}_{k \ell} \,{P}{[}r{]}^\ell{}_j  .
\]
For the right-hand side to be a valid symmetrised projector, note that we must have $a_{rs} = a_{sr}$. We conjecture the following Theorem
\mybenv{Theorem (conjecture)} \textbf{Second level atypicality conditions for $gl_2(n/1)$}:\\
\label{thm:Conjecture}
Let $\delta_{rs}$ be a weight of $V_0(\Lambda_2^*)$. The irreducible Kac module $V(\Lambda)$ contains the $gl(n)$ submodule $V_0(\Lambda+ \delta_{rs})$ iff $\Lambda + \delta_{rs}$  is a dominant integral weight, and $a_{rs}(\Lambda_1,\Lambda_2,\cdots, \Lambda_n) \ne 0$. \hfill 
\\ \mbox{} \myeenv 
For higher levels, an expression equivalent to (\ref{eq:Bdecomp}) can be developed for the appropriate higher rank antisymmetric tensors and their shift projections, leading to an inductive approach to handling the structure of the atypicality conditions governing the presence or absence of the various even irreducible modules. We defer the details to future work.

\end{appendix}
\end{document}